\documentclass[sigconf]{acmart}

\usepackage{multirow}
\AtBeginDocument{%
  \providecommand\BibTeX{{%
    \normalfont B\kern-0.5em{\scshape i\kern-0.25em b}\kern-0.8em\TeX}}}

\setcopyright{acmlicensed}
\copyrightyear{2025} 
\acmYear{2025}
\acmConference[CHI '25]{CHI Conference on Human Factors in Computing Systems}{April 26-May 1, 2025}{Yokohama, Japan}
\acmBooktitle{CHI Conference on Human Factors in Computing Systems (CHI '25), April 26-May 1, 2025, Yokohama, Japan}
\acmDOI{10.1145/3706598.3713100}
\acmISBN{979-8-4007-1394-1/25/04}

\begin{document}

\title{BIT: Battery-free, IC-less and Wireless Smart Textile Interface and Sensing System}

\author{Weiye Xu}
\orcid{0000-0001-5031-8154}
\affiliation{%
  \institution{Tsinghua University}
  \city{Beijing}
  \country{China}
}
\email{xuwy24@mails.tsinghua.edu.cn}

\affiliation{%
  \institution{Simon Fraser University}
  \state{BC}
  \country{Canada}
}
\email{wxa44@sfu.ca}

\author{Tony Li}
\orcid{0009-0000-7275-1637}
\affiliation{%
  \institution{Stony Brook University}
  \city{Stony Brook}
  \state{NY}
  \country{United States}
}
\email{haolili@cs.stonybrook.edu}

\author{Yuntao Wang}
\authornote{Corresponding authors.}
\orcid{0000-0002-4249-8893}
\affiliation{%
  \institution{Tsinghua University}
  \city{Beijing}
  \country{China}
}
\email{yuntaowang@tsinghua.edu.cn}

\author{Xing-Dong Yang}
\orcid{0000-0002-6732-6748}
\affiliation{%
  \institution{Simon Fraser University}
  \state{BC}
  \country{Canada}
}
\email{xingdong_yang@sfu.ca}

\author{Te-Yen Wu}
\authornotemark[1]
\orcid{0000-0003-3977-9093}
\affiliation{%
  \institution{Florida State University}
  \state{FL}
  \country{United States}
}
\email{tw23l@fsu.edu}

\renewcommand{\shortauthors}{Xu, et al.}
\newcommand{\MK}[1]{\textcolor{black}{#1}}
\newcommand{\MG}[1]{\textcolor{black}{#1}}

\begin{abstract}
  The development of smart textile interfaces is hindered by the inclusion of rigid hardware components and batteries within the fabric, which pose challenges in terms of manufacturability, usability, and environmental concerns related to electronic waste. To mitigate these issues, we propose a smart textile interface and its wireless sensing system to eliminate the need for ICs, batteries, and connectors embedded into textiles. 
  Our technique is established on the integration of multi-resonant circuits in smart textile interfaces, and utilizing near-field electromagnetic coupling between two coils to facilitate wireless power transfer and data acquisition from smart textile interface.
  A key aspect of our system is the development of a mathematical model that accurately represents the equivalent circuit of the sensing system. Using this model, we developed a novel algorithm to accurately estimate sensor signals based on changes in system impedance. Through simulation-based experiments and a user study, we demonstrate that our technique effectively supports multiple textile sensors of various types.   
\end{abstract}

\ccsdesc[500]{Human-centered computing~Interaction devices}

\keywords{battery-less, IC-less, wireless, smart textile, sensing interface, relay}

\begin{teaserfigure}
  \includegraphics[width=\textwidth]{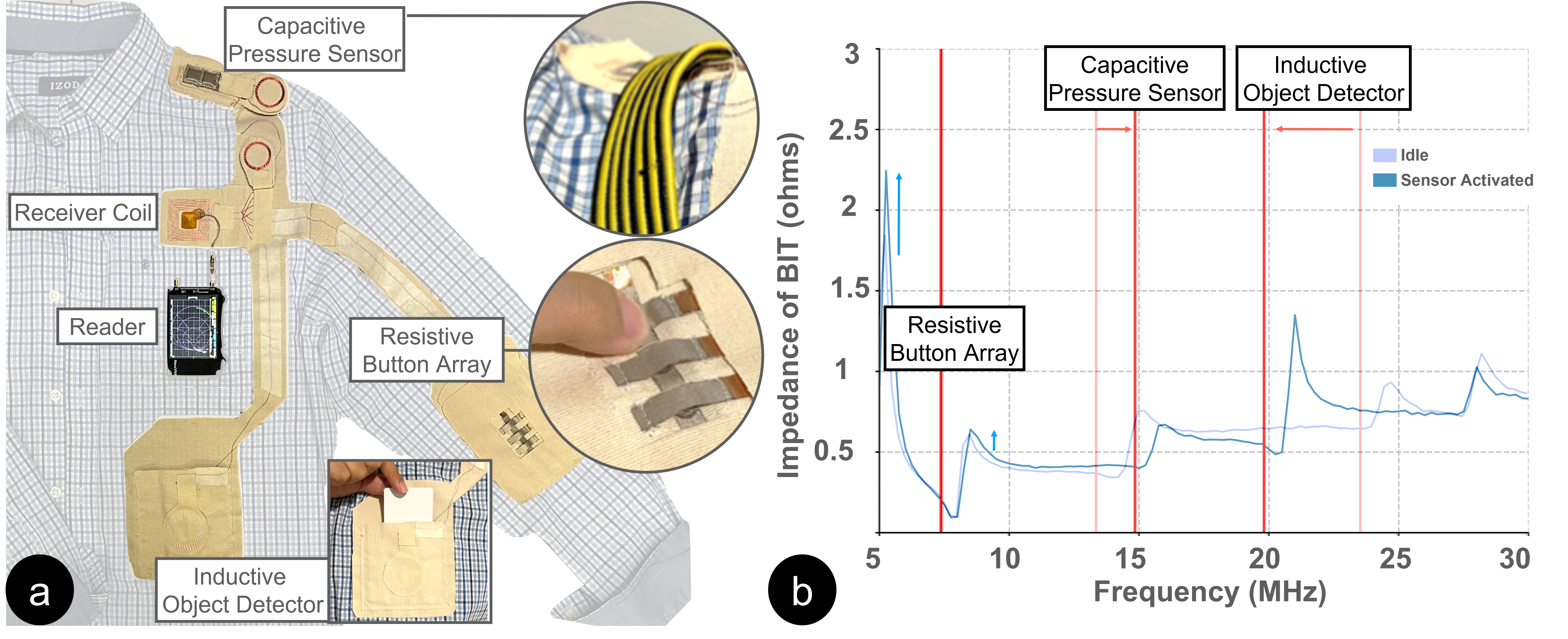}
  \caption{BIT is a smart textile interface that operates without the need for embedding batteries, ICs and connectors into textiles. It features with a receiver coil, enabling wireless coupling and sensor data acquisition with a reader. BIT can be designed to accommodate sensors based on capacitance, resistance, or inductance and support concurrent operations of up to three sensors. (a) For example, a smart shirt was created with BIT, incorporating a resistive button array, a capacitive pressure sensor and an inductive object detector. (b) When interactions occur with the embedded sensors, the impedance spectrum of the smart textile interface changes. This change is wirelessly measured by the reader and can be further interpreted as sensor signals.}
  \Description{This figure demonstrates a overview of a prototype of BIT system, a smart shirt and the impedance plot of BIT. BIT is a smart textile interface that operates without the need for embedding batteries, ICs and connectors into textiles. It features with a receiver coil, enabling wireless coupling and sensor data acquisition with a reader. BIT can be designed to accommodate sensors based on capacitance, resistance, or inductance and support concurrent operations of up to three sensors. (a) For example, a smart shirt was created with BIT, incorporating a resistive button array, a capacitive pressure sensor and an inductive object detector. (b) When interactions occur with the embedded sensors, the impedance spectrum of the smart textile interface changes. This change is wirelessly measured by the reader and can be further interpreted as sensor signals.}
  \label{fig:teaser}
\end{teaserfigure}


\maketitle

\section{Introduction}
\MK{The development of smart textile technology has enabled integration of daily user input into wearable items such as garments, gloves, and bags, offering an alternative to traditional devices like touchscreens. However, implementing smart textiles faces challenges in manufacturability \cite{Berglund2015, Molla2017, Zhu2022}, usability \cite{Chen2019, Seyed2021, Wang2013}, and environmental sustainability \cite{Ossevoort2013, Tat2022, Timmins2009}, primarily due to the embedment of rigid electronic components such as batteries and circuits into textile interfaces. These components compromise comfort, flexibility as well as usability, and contribute to electronic waste when textile products become obsolete.}

To address these issues, various methods have been proposed \cite{Garnier2021, Garnier2023, Hajiaghajani2021, Huang2021, Lin2020, Ye2022, Singlebodycoupledfiber, Flextouch}, among which resonant sensors hold significant promise \cite{Charkhabi2021, Galli2023, Grimes2002, Lee2018, Riistama2010, Salpavaara2018}. \MK{This approach utilizes minimal passive components, typically an LC (inductor-capacitor) resonant circuit consisting of a receiver coil and a capacitive sensor. When the capacitive sensor is pressed, the resonant frequency changes, and this change can be detected through near-field magnetic coupling with a reader. The reader, similar in form factor to a smartphone, can be placed in a pocket, enabling seamless interactions and minimizing the need for embedded electronics in textiles.}


\MK{However, several challenges remain. First, the approach currently focuses solely on capacitive sensing, limiting its applicability to other wearable sensors that rely on inductance and resistance. Additionally, the design does not account for transmission line deformation or body contact, which can introduce fluctuations in capacitance and inductance, affecting sensor readings. Furthermore, misalignments between the reader and interface coils, common in wearable contexts, may reduce accuracy. Lastly, the approach supports only single-sensor operation per reader, limiting its usability, as most users carry only one personal device, such as a smartphone.}

\MK{
In this paper,
we proposed an alternative method. Our approach extends the resonator-based technique by involving N-parallel series RLC (resistor-inductor-capacitor) circuits on the the receiver coil (Figure 1a). Unlike existing approaches, our method supports up to three different types of sensors commonly used in smart garment applications, including resistive, capacitive, and inductive sensors. Additionally, our approach takes into account the influence of the transmission lines and coil misalignment, ensuring more accurate and robust sensor readings. Furthermore, our technique allows for the concurrent operation of up to three sensors of the same or different types. When a user interacts with the interface, such as pressing a capacitive sensor, the interaction causes a change in the system’s impedance (Figure 1b), which can then be wirelessly measured and detected by an external reader.}

To demonstrate the effectiveness of the proposed approach, we developed a proof-of-concept prototype that consists of a smart textile interface, which is primarily composed of sensors and a receiver coil sewed on a textile substrate, and a reader comprising a transmitter coil and a vector network analyzer (NanoVNA) for measuring impedance spectrum of the smart textile interface. The specific design of smart textile interfaces can be customized to support a range of resistive, capacitive, or inductive sensor, depending on the specification of sensors and the number of sensors implemented in the interface. 
When the transmitter coil is aligned with the receiver coil on the interface, the reader wirelessly measures the impedance spectrum of the interface. 
Our system, then, uses an algorithm to analyze the measured impedance spectrum for sensor signal estimation. 
This algorithm was developed based on a mathematical model derived from the equivalent circuit of the system,  accounting for real-world factors such as transmission line effects and coil misalignment, both of which can affect the measured impedance. To enhance the accuracy and efficiency of sensor value estimation, this algorithm incorporates one additional known LC circuit within the interface. Through our simulation-based experiments, we found that our system can reliably capture the sensor signals with an average accuracy over 90\%. Additionally, we conducted a user study to validate the system's performance in real world conditions.  The results demonstrated that the system robustly captured sensor signals generated by user interactions with an augmented shirt, and achieved an overall accuracy of 93\% in classifying user interactions. 



The key contributions of this work include:
\begin{itemize}
    \item An approach that uses the structure of N-parallel series RLC circuits to address the limitation of resonance-based sensors in battery-free, IC-less and wireless smart textile interfaces;
    \item A mathematical model and algorithm that enable our sensing system to accurately estimate sensor values from measured impedance spectrum;
    \item Experimental results demonstrating the effectiveness of the proposed approach.
\end{itemize}

\section{Related Work}
This section provides a brief overview of the existing literature on textile-based input interfaces, textile resonators, and e-textile fabrication and manufacturing. 

\subsection{Textile-based Input Interfaces}

Textile-based input interfaces provide a versatile platform for interacting with electronic devices \cite{Parzer2018, Poupyrev2016, Vogl2017, Wu2020Fabriccio}. These interfaces can be categorized into explicit and implicit inputs. Explicit input interfaces involve direct interactions, such as touch gestures like tapping and swiping \cite{Aigner2021, Heller2014, Ku2020, Parzer2018, Parzer2017, Poupyrev2016}, as well as deformation gestures like stretching \cite{Vogl2017}, bending \cite{Glauser2019, Parzer2017}, and squeezing \cite{Olwal2018, Olwal2020}, and non-contact gestures like waving \cite{Wu2020Fabriccio}. In contrast, implicit input does not require specific actions; instead, it gathers information through monitoring user activities and contexts. Textile-based implicit input interfaces have found practical applications in areas such as activity tracking \cite{Cheng2013, Glauser2019, Lee2009, Liu2019}, health monitoring \cite{Ahsan2022, Charkhabi2021, Lee2009, Majumder2017, Xu2013}, and contextual interactions \cite{Gong2019, Wu2020Capacitivo, Wu2021}. One example is pressure-sensitive textile cushions that detect seated posture for ergonomic adjustments \cite{Xu2013}. Recent research by Wu et al. \cite{Wu2020Capacitivo, Wu2021} has shown that smart textiles can recognize objects in contact, enabling fine-grained activity recognition and contextual interactions. Additionally, smart textiles are widely used for tracking body movements in physical activities \cite{Cheng2013, Glauser2019, Lee2009, Liu2019}, providing valuable data for health monitoring \cite{Ahsan2022, Majumder2017}.

\MK{Despite many applications, most of the existing work necessitates the incorporation of rigid hardware components and batteries into textile interfaces. 
Our research explores an alternative approach that eliminates integrated circuits, batteries, and connectors by using a magnetic resonant coupling technique. This approach relies on an external reader, like a smartphone in the user’s pocket, to wirelessly drive and read the smart textile interface, reducing embedded electronics and enabling more flexible and sustainable smart textile interfaces.}

\subsection{Textile Resonators}

Resonators are circuits that exhibit electrical resonance at specific frequencies. Over time, they have taken on various forms and have found applications in domains such as wireless communication \cite{Garnier2021, Garnier2023, Ye2022}, power transfer \cite{Seo2016, Wagih2020}, and  sensing \cite{Charkhabi2021, Galli2023, Grimes2002, Lee2018, Riistama2010, Salpavaara2018}. Our research is related to two types of resonators: relay resonators and resonant sensors.

\MK{Relay resonators use resonant coupling to wirelessly transfer power or communication signals, with applications in wireless charging \cite{Saha2018, Wagih2020, Zhu2015}, sensor networks \cite{Lee2009, Lin2020}, and on-body communication \cite{Garnier2021, Garnier2023, Ye2022}. These systems typically include an inductive coil and capacitor, often made with textile materials \cite{Garnier2021}, and use a magnetic field to induce current, which can power devices or facilitate communication. In on-body networks, the relay resonator can support signal coverage up to 1 meter by adjusting the capacitor for body capacitive coupling \cite{Ye2022}.  This extended coverage allows for greater flexibility in wearable applications.}

Another relevant technology in the field is the resonant sensor, specifically the inductively coupled resonance sensor.  These sensors can be constructed using textile materials, as they rely on simple and passive components such as capacitors and inductors \cite{Charkhabi2021, Galli2023, Grimes2002, Lee2018, Riistama2010, Salpavaara2018, Sun2021}. Resonant sensors utilize the principles of resonance to detect and measure physical quantities such as temperature \cite{Ibanez-Labiano2020}, pressure \cite{Sun2021}, or fluid conductivity \cite{Charkhabi2021}. 
However, these sensors are currently limited in type, typically functioning based on capacitance. In addition, they did not account for the influence of transmission lines and coil misalignment, which can significantly affect the sensor readings. Furthermore, none of the existing systems can support multiple sensors to simultaneously operate. This makes each sensor individually connected to a coil for wireless power transfer and data acquisition, limiting the usability of the approach. 

To address these issues, we designed and developed a battery-free and IC-less smart textile interface that can support a variety of types of sensors while functioning under the influence of transmission lines, and slight coil misalignment. It can also support the concurrent operation of up to three sensors, broadening the applicability of this interface. We analyzed the circuit model and developed a mathematical model and algorithm to enable the reader to accurately interpret sensor data from multiple sensors of various types, allowing the system to effectively capture a wide range of user interactions through textile surfaces or garments. 

\subsection{E-Textile Fabrication and Manufacturing}
\MK{The fabrication of electronic textiles (e-textiles) is challenging due to the complexity of embedding electronics into fabrics. To address this, various toolkits have been developed. Notably, Lilypad \cite{Buechley2008, Buechley2010} enables hobbyists to incorporate electronics into garments, while MakerWear \cite{Kazemitabaar2015, Kazemitabaar2017} offers a modular approach for beginners. These toolkits have inspired the use of modular electronics and block-based programming in an avant-garde runway environment \cite{Seyed2021}. Other research initiatives, such as Teeboard \cite{Ngai2009}, I*CATch \cite{Ngai2010}, and E-broidery \cite{Post2000}, instead explore integrating e-textiles through embroidery machines \cite{Aigner2021, Aigner2020, Blecha, Hamdan2018}. Additionally, Klamka et al. developed an iron-on toolkit for easy bonding of e-textiles to fabric \cite{Klamka2020}. }

\MK{In addition to these tools for small-scale fabrication, researchers have also explored the large-scale manufacturing of e-textiles. For instance, Molla et al. \cite{Molla2017} studied creating e-textile circuits on a larger scale using reflow soldering and conductive threads, demonstrating that small 2-pin SMD components like LEDs can withstand typical washing and wear. Zhu and Kao \cite{Zhu2022} identified four key challenges in large-scale e-textile manufacturing: the lack of production standards, disconnects between apparel and hardware manufacturing, cost disparities, and limited production-capable solutions. Addressing these challenges is essential for scaling up e-textile production and meeting industry demands.}

Building on prior research, our work aims to eliminate the need for e-textile interfaces to be integrated with rigid hardware components such as batteries, and connectors. This approach has the potential to significantly simplify the fabrication and manufacturing process of e-textiles as well as make e-textiles more sustainable and eco-friendlier, lowering the environmental impacts brought by the realization of ubiquitous computing.

\section{BATTERY-FREE, IC-LESS AND WIRELESS SMART TEXTILE INTERFACE}
Our goal is to develop a smart textile interface that operates without the inclusion of batteries and ICs within textiles, while retaining the ability to support various types of textile sensors. To achieve this goal, our proposed approach is based on resonant sensors, which utilize the characteristics of resonant circuits to wirelessly reflect the sensor signals to an external reader, eliminating the need for embedding batteries and ICs into textiles.
This section discusses the operating principle, and provides further understanding in the electrical behavior of this interface.

\subsection{Operating Principle}

\begin{figure*}[htbp]
    \centering
    \includegraphics[width=1.0\linewidth]{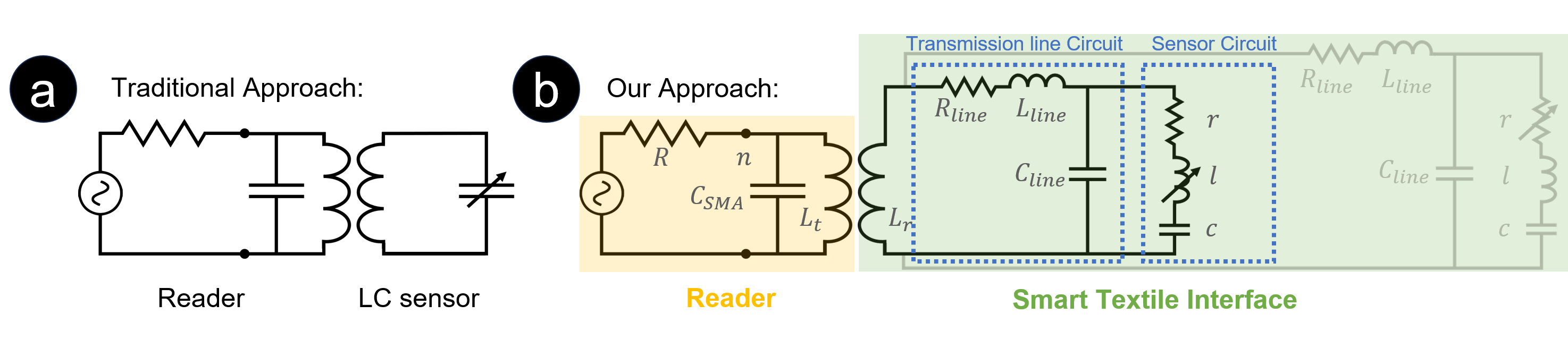}
    \caption{(a) The equivalent circuit of a traditional resonant sensor system. (b) The equivalent circuit of our entire system, which consists of a reader circuit and a smart textile interface circuit. The interface includes a receiver coil ($L_r$) connected in parallel with multiple transmission and sensor circuits. Each transmission line circuit contains parasitic resistance ($R_{line}$), inductance ($L_{line}$), and capacitance ($C_{line}$), connected to a sensor circuit with a resistor (r), capacitor (c), and inductor (l) in series. The reader circuit includes a voltage exciter, internal load (R), and transmitter coil ($L_t$), with parasitic capacitance ($C_{SMA}$). Impedance is measured by capturing voltage at point (n) and applying the voltage divider rule. }
    \Description{
    This figure demonstrates the circuit model of BIT, compared with traditional LC sensor.
    (a) The equivalent circuit of a traditional resonant sensor system. (b) The equivalent circuit of our entire system, which consists of a reader circuit and a smart textile interface circuit. The interface includes a receiver coil ($L_r$) connected in parallel with multiple transmission and sensor circuits. Each transmission line circuit contains parasitic resistance ($R_{line}$), inductance ($L_{line}$), and capacitance ($C_{line}$), connected to a sensor circuit with a resistor (r), capacitor (c), and inductor (l) in series. The reader circuit includes a voltage exciter, internal load (R), and transmitter coil ($L_t$), with parasitic capacitance ($C_{SMA}$). Impedance is measured by capturing voltage at point (n) and applying the voltage divider rule. 
    }
    \label{fig:eq_circuit}
\end{figure*}

\MK{
A traditional resonant sensor comprises an LC resonant circuit, the resonant frequency of which can be wirelessly monitored using a reader through magnetic coupling. Its equivalent circuit is described in Figure \ref{fig:eq_circuit}a. To support more common sensor types, including capacitive, resistive, and inductive, we replaced the capacitor (C) in the typical LC circuit by a series RLC resonant circuit, as shown in Figure \ref{fig:eq_circuit}b. This resonant circuit, referred as a sensor circuit later, consists of a resistor, inductor, and capacitor, designed to operate within a specific frequency range. Different sensors can be supported by replacing the corresponding components. For example, replacing the inductor with a coil forms an inductive sensor for detecting metallic objects. To enable concurrent operation of multiple sensors, we connected multiple sensor circuits in parallel, each designed to operate at distinct frequency ranges. This allows our smart textile interface to detect inputs from multiple sensors with minimal passive components.}

\MK{During operation, the oscillating magnetic field produced by the reader's transmitter coil induces a current within the smart textile interface through near-field magnetic coupling with the receiver coil. User interaction with the sensor induces a change in the value of the corresponding component within the RLC resonant circuit (e.g., inductance variations due to metallic objects), causing a shift in the impedance spectrum of the smart textile interface. Consequently, this affects the current circulating within the transmitter coil, ultimately resulting in changes in the impedance spectrum within the reader circuit. Our system measured these changes and used an algorithm to extract sensor values for the detection of user input.}

\MK{Although this approach seems feasible, numerous factors actually influence the impedance spectrum of the smart textile interface, complicating wireless sensor readings. These factors include the design of the sensor circuits, the parasitic capacitance and inductance of transmission lines, and the misalignment between the transmitter and receiver coils. To address these challenges, we first gained a deep understanding of the circuits model within the smart textile interface and developed corresponding solutions to them. }

\subsection{Equivalent Circuit Model}
\label{sec_equivalent_circuit}
\MK{The equivalent circuit model is a simplified representation of the system's electrical behavior. Different versions exist, balancing accuracy and simplicity based on the level of detail needed. For example, in some models \cite{Garnier2023, Garnier2021}, transmission line behavior could be simplified to just parasitic capacitance and resistance, neglecting the effects of parasitic inductance. For our system, our goal was to develop an equivalent circuit model that is accurate enough to capture key behaviors and simple enough for the analysis of our system. Based on our pilot study, we designed our model as shown in Figure \ref{fig:eq_circuit}b, mainly following prior research on relay resonators \cite{Garnier2021}.}
Using this model, we have formulated the following equations to describe the impedance of the entire system, based on the literature on wireless power transfer systems \cite{Seo2016}:

\begin{gather}
    \label{eq_z}
    Z(f) = \frac{1}{\frac{1}{(2\pi fL_t)j-\frac{Z_M(f)^2}{Z_s(f)}}+2\pi fC_{SMA}j} \\
    \label{eq_zm}
    Z_M(f) = (2\pi f)k\sqrt{(L_t L_r)}j \\
    \label{eq_zs}
    Z_S(f) = \frac{1}{\sum^n_1 \frac{1}{Z_i(f)}}+(2\pi fL_r)j \\
    \label{eq_zi}
    Z_i(f) =\frac{1}{\frac{1}{(2\pi fl_i-\frac{1}{2\pi fc_i})j + r_i}+2\pi fC_{line_i}j}+2\pi fL_{line_i}j+R_{line_i}
\end{gather}

where \MK{$Z$ represents the total measured impedance, $f$ is the operating frequency, $Z_M$ is the impedance of mutual inductance, $k$ is the coupling factor between the transmitting and receiving coils, $Z_S$ is the impedance of the smart textile interface, $Z_i$ is the impedance of the $i^{th}$ sensor circuit connected through a transmission line, $n$ is the number of resonant circuits embedded in the interface, and $j$ is the imaginary unit.}

\subsubsection{Circuit Model Validation}
\label{sec_model_validation}

\begin{figure*}[htbp]
    \centering
    \includegraphics[width=1\linewidth]{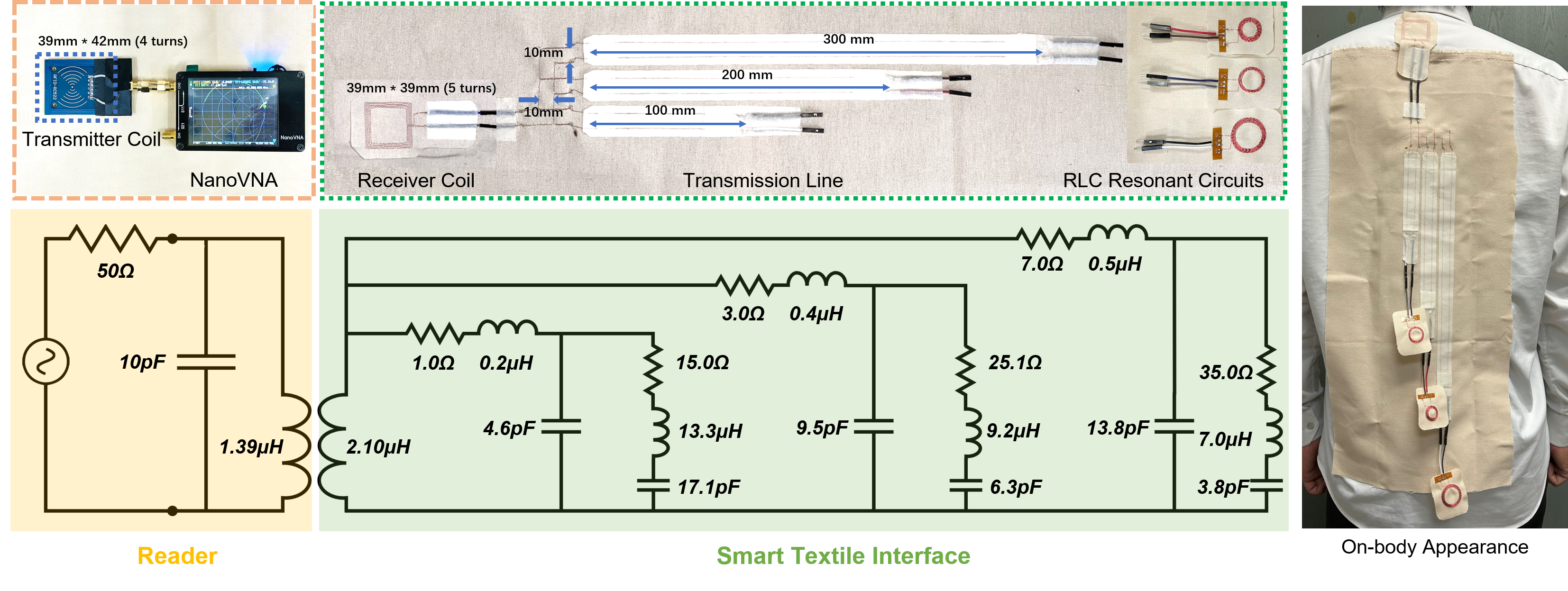}
    \caption{\MK{The prototype used in our experiment to validate the accuracy of our equivalent circuit model. The reader was constructed using a NanoVNA \cite{NanoVNA} connecting to a standard NFC transmitter coil (39mm by 42mm with 4 turns) \cite{MFRC-522coil} through a SubMiniature version A (SMA) connector \cite{SMAconnector}. For the smart textile interface, each sensor circuits consisted of a resistor, an inductor and a capacitor, connecting to an embroidered receiver coil (39mm by 39mm with 5 turns) in parallel via transmission line with lengths of 100mm, 200mm, 300mm and same gap of 10mm. Detailed physical parameters and electrical attributes of components are illustrated in the figure. The coupling factor of the two coils was measured to be around 0.53 using a 2-port VNA to measure mutual inductance \cite{Jeon2019}. The fabrication process is the same as described in Section \ref{sec_implementation_of_smart_textile_interface}.}}
    \label{fig:fig3-Apparatus}
    \Description{
    This figure demonstrates a prototype and the according circuit model with detailed value of r, l and c components.
    The prototype used in our experiment to validate the accuracy of our equivalent circuit model. The reader was constructed using a NanoVNA \cite{NanoVNA} connecting to a standard NFC transmitter coil (39mm by 42mm with 4 turns) \cite{MFRC-522coil} through a SubMiniature version A (SMA) connector \cite{SMAconnector}. For the smart textile interface, each sensor circuits consisted of a resistor, an inductor and a capacitor, connecting to an embroidered receiver coil (39mm by 39mm with 5 turns) in parallel via transmission line with lengths of 100mm, 200mm, 300mm and same gap of 10mm. Detailed physical parameters and electrical attributes of components are illustrated in the figure. The coupling factor of the two coils was measured to be around 0.53 using a 2-port VNA to measure mutual inductance \cite{Jeon2019}. The fabrication process is the same as described in Section \ref{sec_implementation_of_smart_textile_interface}.
    }
\end{figure*}
\MK{
To validate the effectiveness of the circuit model, we implemented a hardware prototype and compared the impedance spectrum generated by the model with the actual spectrum from the prototype. The prototype, consisting of a reader made from NanoVNA and a smart textile interface with three sensor circuits (Figure \ref{fig:fig3-Apparatus}), was affixed to the back of a collared shirt worn by a 23-year-old male volunteer. During data collection, the reader measured the S11 reflection coefficient from 1 MHz to 40 MHz to retrieve the ground-truth impedance spectrum. S11 quantifies the portion of a wave reflected by impedance discontinuities and is easier to measure accurately with our reader than the impedance spectrum itself. Similar to impedance, S11 also has real and imaginary components, which can be measured separately using NanoVNA.  On the other hand, with the parameters described in Figure \ref{fig:fig3-Apparatus}, we calculated an estimation of the impedance spectrum using our model (Eq. 1-4). Then, we derived the S11 values using the following formula:}

\begin{gather}
    S11 = \frac{Z - 50}{Z + 50} 
\end{gather}

where 50$\Omega$ is the standardized internal impedance of NanoVNA.

\begin{figure}
    \centering
    \includegraphics[width=1\linewidth]{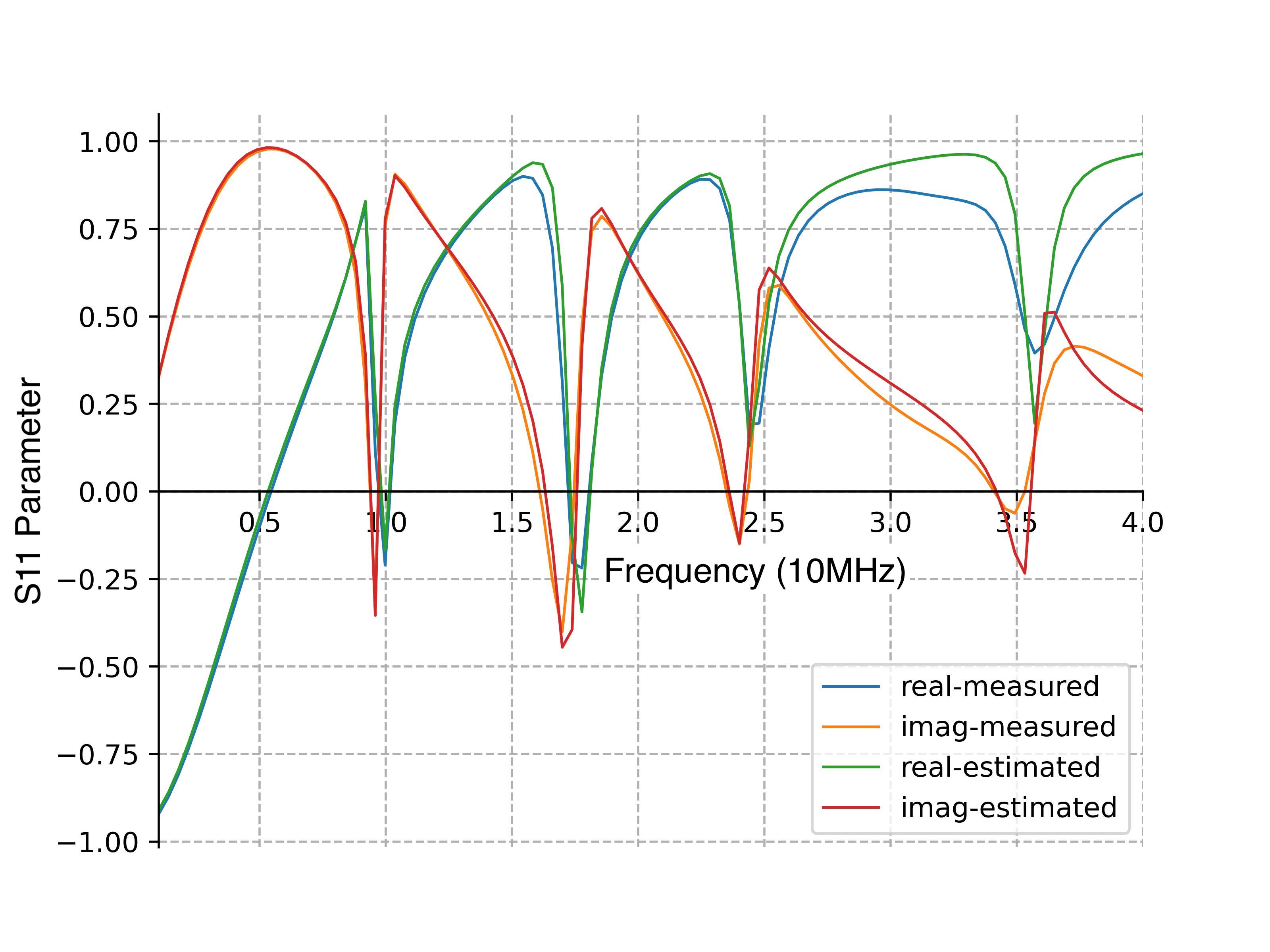}
    \caption{The ground truth S11 values (blue and orange) and the predicted results using our model (green and red).}
    \Description{
    This figure demonstrate the S11 plot of the prototype, including the ground truth S11 values (blue and orange) and the predicted results using our model (green and red).
    }
    \label{fig:fig3-Results}
\end{figure}

Finally, we compared our estimated S11 values with the measured ground-truth S11 values.
Figure \ref{fig:fig3-Results} illustrates the comparison. Overall, our estimation aligns relatively well with the ground truth values ($R^2 = 0.96$), despite small discrepancies in the frequency range higher 30M Hz. We suspect that this may be attributed to the capacitive coupling between two coils. To mitigate this effect, we restricted the operating frequency of the system within the range of 1M Hz and 30M Hz.

\subsubsection{Discussion}
\MK{
From the validated circuit model, we derived several key insights. First, each resonant circuit should incorporate only one type of sensor—resistive, capacitive, or inductive—while keeping other components fixed, as recommended by Eq. 4. This reduces unknown variables, increasing model accuracy and solving speed when the operating frequency is within the circuit's resonant range.
Second, the coupling factor ($k$), as shown in Eq. 1 and Eq. 2, is crucial in determining how user-induced impedance changes are observed in the reader’s impedance spectrum. A higher $k$ leads to more pronounced impedance changes, but consistency in $k$ is necessary to avoid inconsistencies in measured impedance spectrum. $k$ is influenced by coil design and alignment, with misalignment being inevitable in real-world conditions. Thus, transmitter and receiver coils should be designed for a high coupling factor and tolerance to misalignment, even when placed in a small pocket. Additionally, high coil inductance increases total impedance at high frequencies, reducing sensor-induced impedance changes, so a trade-off must be made during coil design.
Finally, Eq. 4 suggests that transmission lines should be designed to minimize variation in capacitance ($C_{line}$) and inductance ($L_{line}$) and to keep $C_{line}$ low. High variation complicates impedance spectrum changes, reducing reading accuracy, while high $C_{line}$ can short-circuit high frequencies, diminishing sensor circuit changes. Based on these insights, our next step was to optimize coil design and transmission line design to enhance sensing performance.
}

\subsection{Coil Design}
\label{sec_coil_design}
\MK{To ensure consistent performance of sensor value estimation in the system, we tested various transmitter and receiver coil designs to maintain a high and consistent coupling factor ($k$) in both aligned and misaligned conditions. We selected a rectangular coil design to maximize coverage areas. We explored 9 combinations of 3 transmitter and 3 receiver coil types, with transmitter coils fabricated on Flexible Printed Circuit Boards (similar to phone NFC coils), and receiver coils fabricated on textile substrates via method same as section \ref{sec_implementation_of_smart_textile_interface}. We tested the coupling factor ($k$) by aligning each coil pair at three positions: perfect alignment (0mm), slight misalignment (5mm), and moderate misalignment (10mm).  Using a two-port VNA connecting to transmitter and receiver coil, we measured mutual inductance to determine the coupling factor \cite{Jeon2019}. The results and coil design parameters are presented in Table \ref{table:coil_designs}. }

Our results indicated that larger transmitter coils produced higher $k$ values when aligned but performed poorly under misalignment. We eliminated coil designs with $k$ values below 0.2 under misalignment. Among the remaining coil designs, we selected the transmitter coil with the width of 10mm , and the receiver coil with the width of 40mm for our subsequent studies and implementation. This choice was made because they maintained consistent $k$ higher than 0.25 across different conditions. Additionally, their inductance were low, resulting in smaller reactance in the high frequency range, maximizing the prominence of impedance changes caused by the sensor circuits. Furthermore, the smaller transmitter coil is better suited for integration into compact devices, such as smartwatches. However, it is important to note that the findings of our study are not the optimal result. By exploring a more diverse range of transmitter and receiver coil types, a better coil design pair may be discovered.

\begin{table*}[htbp]
\centering
\begin{tabular}{|l|c|c|c|}
\hline
 & \multicolumn{3}{c|}{\textbf{Receiver Coil}} \\ \cline{2-4} 
 \textbf{Transmitter Coil}
 & \begin{tabular}[c]{@{}c@{}} \textbf{Outer: 20mm} \\ \textbf{Inner:0mm} \\ \textbf{Inductance: 1.21 $\mu$H}\end{tabular} & \begin{tabular}[c]{@{}c@{}} \textbf{Outer: 30mm} \\ \textbf{Inner:10mm} \\ \textbf{Inductance: 2.93 $\mu$H} \end{tabular} & \begin{tabular}[c]{@{}c@{}} \textbf{Outer: 40mm} \\ \textbf{Inner:20mm} \\ \textbf{Inductance: 4.54 $\mu$H} \end{tabular} \\ \hline
 
 \begin{tabular}[c]{@{}c@{}} Outer: 20mm, Inner:0mm \\ Inductance: 0.60 $\mu$H\end{tabular} & \begin{tabular}[c]{@{}c@{}} k (perfect alignment): 0.35\\ k (5mm misalignment): 0.21 \\ k (10mm misalignment): 0.10\end{tabular} & \begin{tabular}[c]{@{}c@{}} k (perfect): 0.29\\ k (5mm): 0.28 \\ k (10mm): 0.24\end{tabular}  & \begin{tabular}[c]{@{}c@{}} k (perfect): 0.25\\ k (5mm): 0.29 \\ k (10mm): 0.29\end{tabular}
\\ \hline

\begin{tabular}[c]{@{}c@{}} Outer: 30mm, Inner:0mm \\ Inductance: 2.31 $\mu$H\end{tabular} & \begin{tabular}[c]{@{}c@{}} k (perfect): 0.71\\ k (5mm): 0.35 \\ k (10mm): 0.13\end{tabular} & \begin{tabular}[c]{@{}c@{}} k (perfect): 0.60\\ k (5mm): 0.44 \\ k (10mm ): 0.16\end{tabular}  & \begin{tabular}[c]{@{}c@{}} k (perfect): 0.33\\ k (5mm ): 0.35 \\ k (10mm ): 0.27\end{tabular}
\\ \hline

\begin{tabular}[c]{@{}c@{}} Outer: 40mm, Inner:0mm \\ Inductance: 7.36$\mu$H\end{tabular} & \begin{tabular}[c]{@{}c@{}} k (perfect): 0.64\\ k (5mm ): 0.44 \\ k (10mm): 0.12\end{tabular} & \begin{tabular}[c]{@{}c@{}} k (perfect): 0.79\\ k (5mm): 0.51 \\ k (10mm): 0.17\end{tabular}  & \begin{tabular}[c]{@{}c@{}} k (perfect): 0.64\\ k (5mm): 0.52 \\ k (10mm): 0.26\end{tabular}
\\ \hline
\end{tabular}
\caption{The dimensions, inductance, and coupling factor (k) of various coil designs.}
\Description{This table illustrates the dimensions, inductance, and coupling factor (k) of various coil designs.}
\label{table:coil_designs}
\end{table*}

\subsection{Transmission Line Design}
\label{sec_transmission_line_design}
\MK{Another critical design factor in the smart textile interface is the transmission line design. Our goals were twofold: minimize variation in parasitic capacitance and inductance, especially in wearable contexts with potential deformation, and reduce parasitic capacitance to free up high-frequency spectrum for more sensor circuits. We explored four transmission line designs: 10mm-spaced, 5mm-spaced, 2.5mm-spaced parallel lines, and twisted lines. These were chosen based on trade-offs between spacing and parasitic effects. Wider spacing, like the 10mm design, reduced capacitance but increased inductance and susceptibility to external interference from the human body and textile deformation. Narrower spacing, such as the twisted design, reduced external interference but increased parasitic capacitance. 
To test how parasitic capacitance and inductance were affected by external influences, we initially measured these values using an LCR meter \cite{DE5000LCRMeter}, then subjected the lines to three conditions: 90-degree bending (simulating typical deformations), 180-degree folding (extreme deformations), and contact with the human body. After each manipulation, we remeasured capacitance and inductance.}

\MK{The results are shown in Table \ref{table:transmission_designs}. We found that parallel transmission lines were significantly affected by the human body, causing notable capacitance changes. Extreme deformation also led to their inductance variations of up to 12\%, 10\%, and 5\% for 10mm, 5mm, and 2.5mm-spaced designs, respectively. In contrast, the twisted transmission line design maintained consistent inductance and capacitance, despite having a higher capacitance ($58pF$). In this work, we selected the twisted design for further study due to its simplicity, though the 2.5mm-spaced parallel design is also a viable option, as its inductance variations were smaller and capacitance variations can be potentially addressed by our sensor value estimation algorithm.}

\begin{table*}[htbp]
\centering
\begin{tabular}{|l|c|c|c|c|}
\hline
 \textbf{}
 & \begin{tabular}[c]{@{}c@{}} \textbf{10mm-spaced} \\ \textbf{transmission lines}\\ \includegraphics[width=0.1\textwidth, height=10mm]{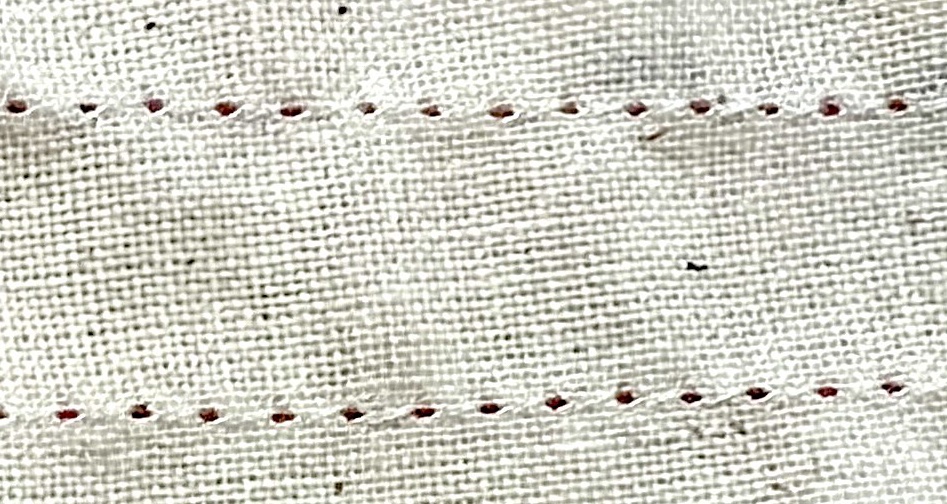}\end{tabular} & \begin{tabular}[c]{@{}c@{}} \textbf{5mm-spaced} \\ \textbf{transmission lines} \\ \includegraphics[width=0.1\textwidth, height=10mm]{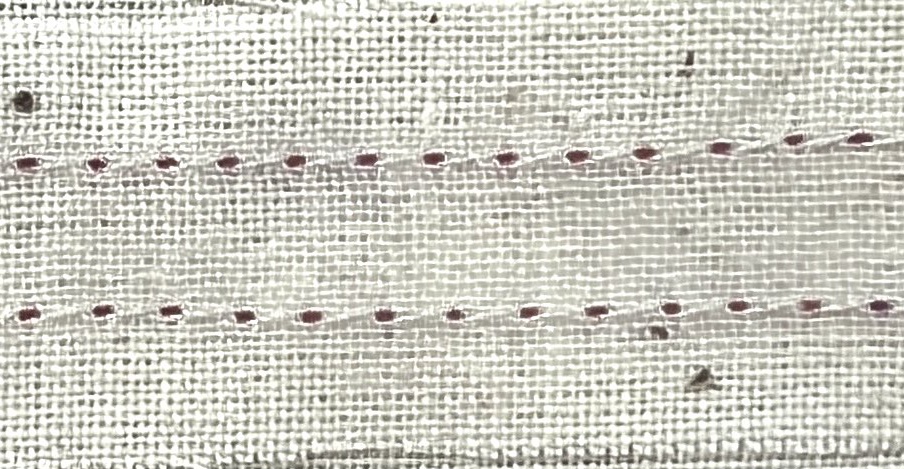}\end{tabular} & \begin{tabular}[c]{@{}c@{}} \textbf{2.5mm-spaced} \\ \textbf{transmission lines} \\ \includegraphics[width=0.1\textwidth, height=10mm]{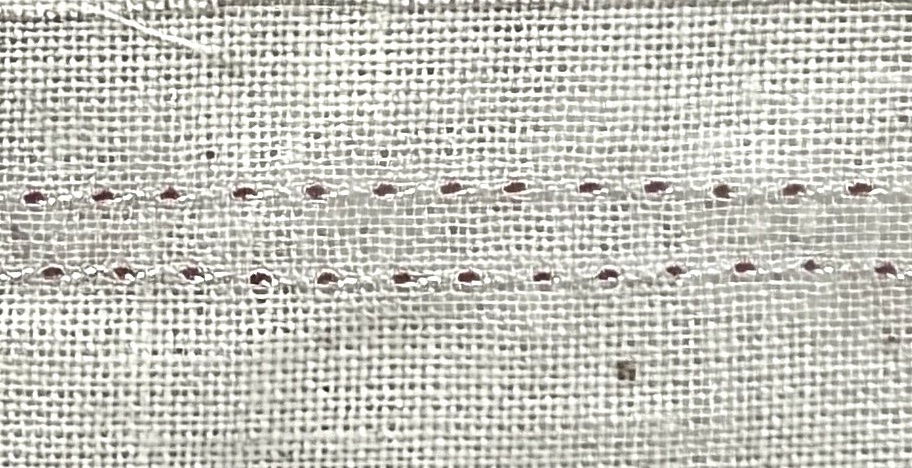}\end{tabular} & \begin{tabular}[c]{@{}c@{}} \textbf{twisted} \\ \textbf{transmission lines}\\ \includegraphics[width=0.1\textwidth, height=10mm]{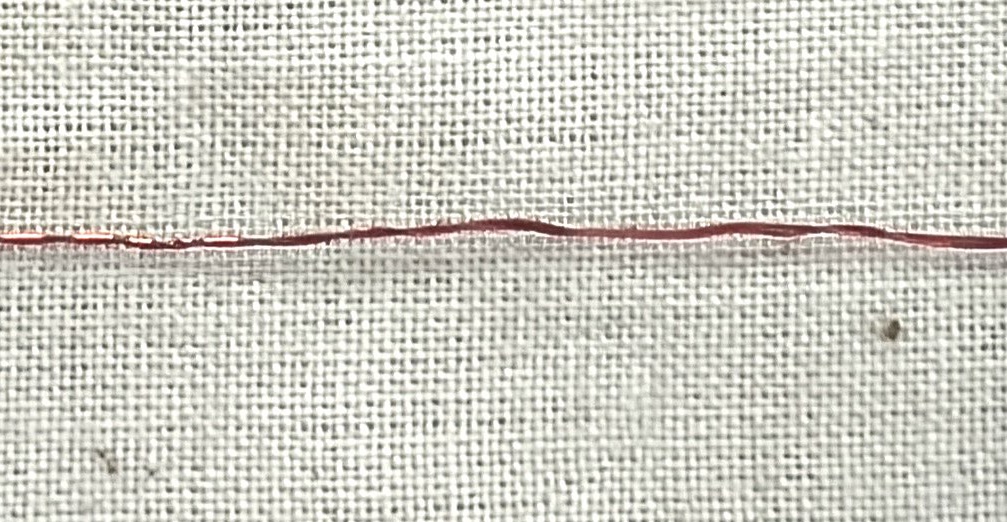} \end{tabular} \\ \hline
 
 \begin{tabular}[c]{@{}c@{}} Straight \\ \includegraphics[width=0.2\textwidth, height=10mm]{Sections/tl_figures/25mm.jpeg} \end{tabular} 
 & \begin{tabular}[c]{@{}c@{}} \MK{capacitance}: 4.76pF \\ inductance:1.22$\mu$H\end{tabular} 
 & \begin{tabular}[c]{@{}c@{}} \MK{capacitance}: 6.19pF \\ inductance:1.15$\mu$H\end{tabular} 
 & \begin{tabular}[c]{@{}c@{}} \MK{capacitance}: 7.5pF \\ inductance:0.95$\mu$H\end{tabular} 
 & \begin{tabular}[c]{@{}c@{}} \MK{capacitance}: 58.33pF \\ inductance:0.21$\mu$H\end{tabular}
\\ \hline

\begin{tabular}[c]{@{}c@{}} Bending at 90 degrees \\ \includegraphics[width=0.2\textwidth, height=10mm]{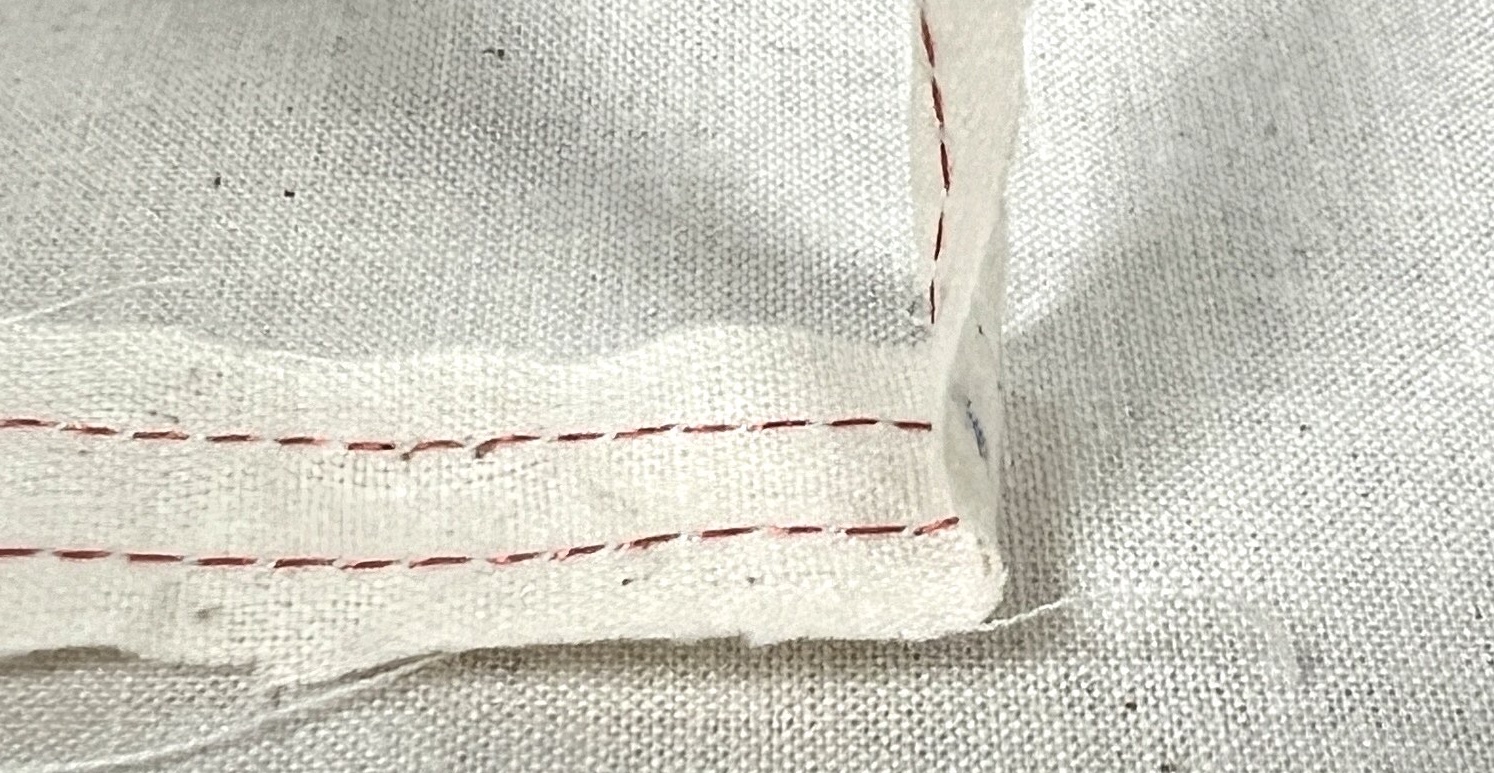} \end{tabular} 
 & \begin{tabular}[c]{@{}c@{}} \MK{capacitance}: 4.47pF \\ inductance:1.23$\mu$H\end{tabular} 
 & \begin{tabular}[c]{@{}c@{}} \MK{capacitance}: 6.10pF \\ inductance:1.15$\mu$H\end{tabular} 
 & \begin{tabular}[c]{@{}c@{}} \MK{capacitance}: 7.45pF \\ inductance:0.96$\mu$H\end{tabular} 
 & \begin{tabular}[c]{@{}c@{}} \MK{capacitance}: 58.06pF \\ inductance:0.21$\mu$H\end{tabular}
\\ \hline

\begin{tabular}[c]{@{}c@{}} Folding at 180 degrees \\ \includegraphics[width=0.2\textwidth, height=10mm]{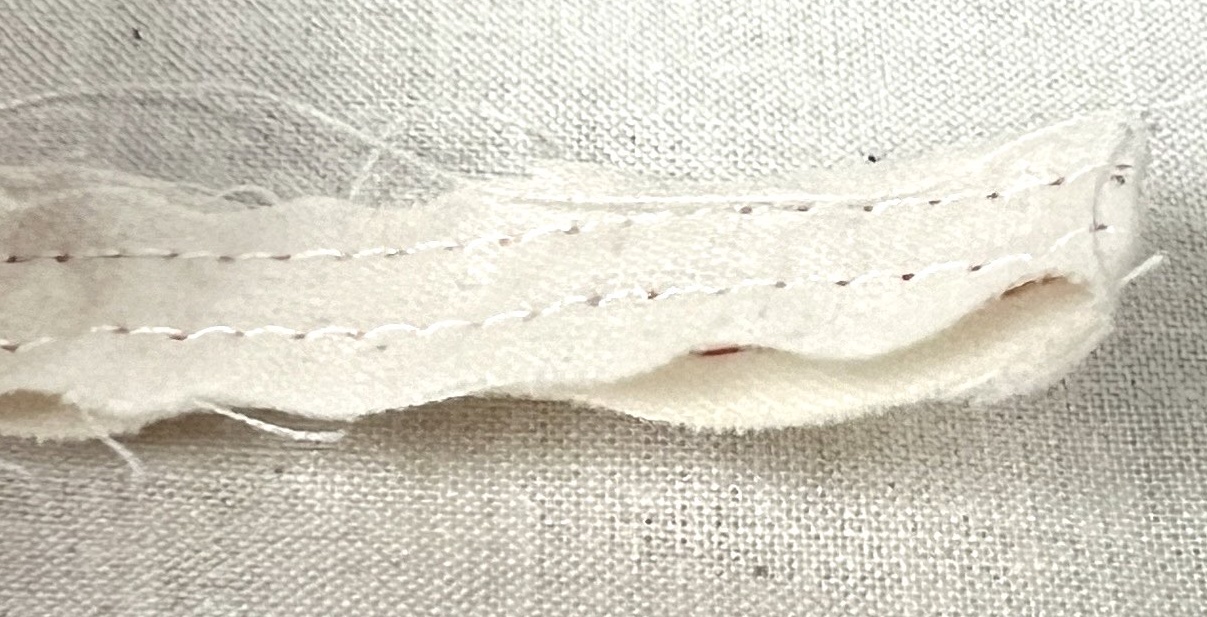} \end{tabular} 
 & \begin{tabular}[c]{@{}c@{}} \MK{capacitance}: 4.97pF \\ inductance:1.07$\mu$H\end{tabular} 
 & \begin{tabular}[c]{@{}c@{}} \MK{capacitance}: 6.42pF \\ inductance:1.03$\mu$H\end{tabular} 
 & \begin{tabular}[c]{@{}c@{}} \MK{capacitance}: 7.67pF \\ inductance:0.89$\mu$H\end{tabular} 
 & \begin{tabular}[c]{@{}c@{}} \MK{capacitance}: 57.64pF \\ inductance:0.21$\mu$H\end{tabular}
\\ \hline

\begin{tabular}[c]{@{}c@{}} On-Body \\ \includegraphics[width=0.2\textwidth, height=10mm]{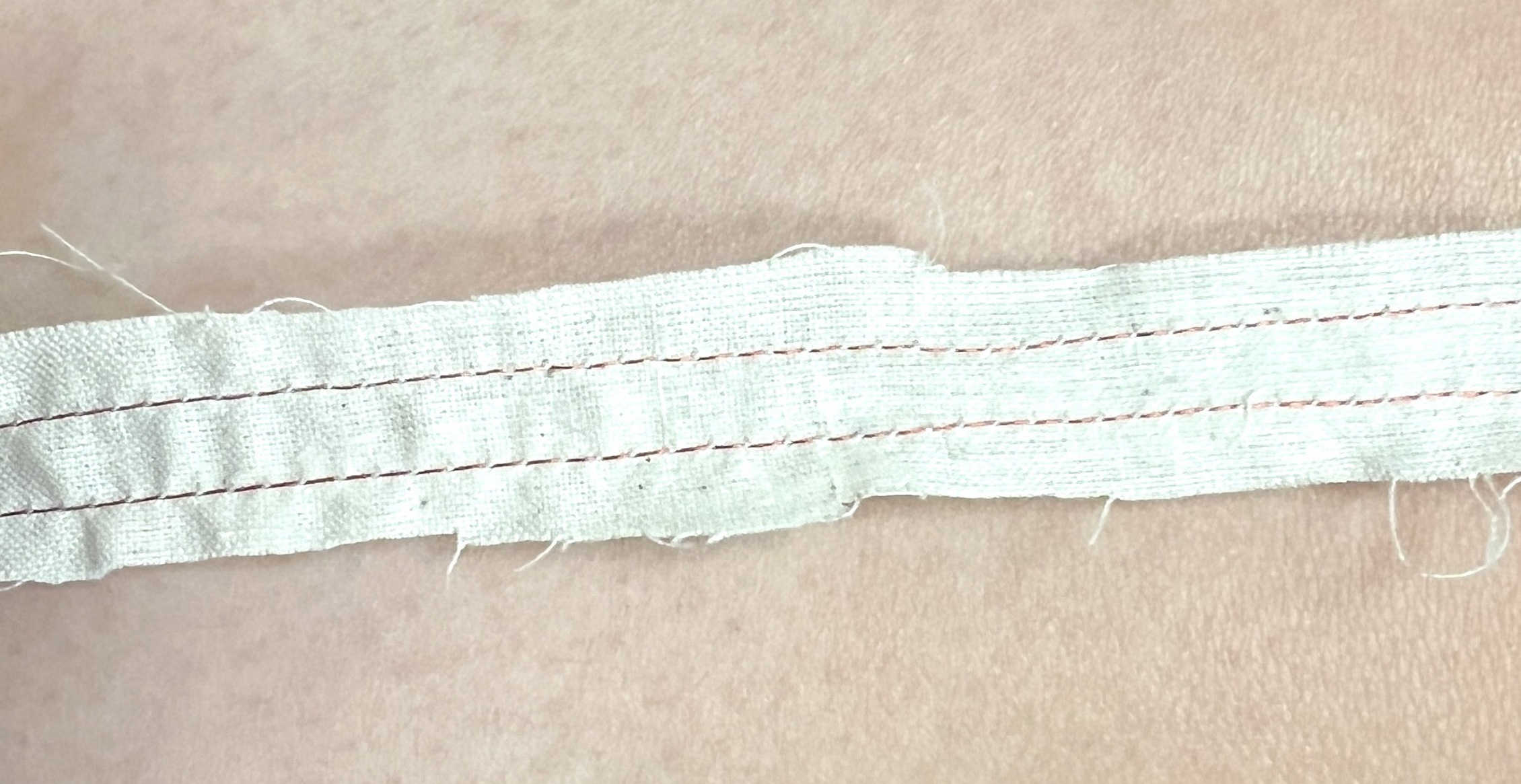} \end{tabular} 
 & \begin{tabular}[c]{@{}c@{}} \MK{capacitance}: 30.53pF \\ inductance:1.23$\mu$H\end{tabular} 
 & \begin{tabular}[c]{@{}c@{}} \MK{capacitance}: 34.26pF \\ inductance:1.15$\mu$H\end{tabular} 
 & \begin{tabular}[c]{@{}c@{}} \MK{capacitance}: 37.82pF \\ inductance:0.95$\mu$H\end{tabular} 
 & \begin{tabular}[c]{@{}c@{}} \MK{capacitance}: 59.62pF \\ inductance:0.22$\mu$H\end{tabular}
\\ \hline
\end{tabular}
\caption{The capacitance and inductance of various transmission line designs under different conditions.}
\Description{This table illustrates the capacitance and inductance of various transmission line designs under different conditions.}
\label{table:transmission_designs}
\end{table*}

\begin{figure}[htbp]
    \centering
    \includegraphics[width=1\linewidth]{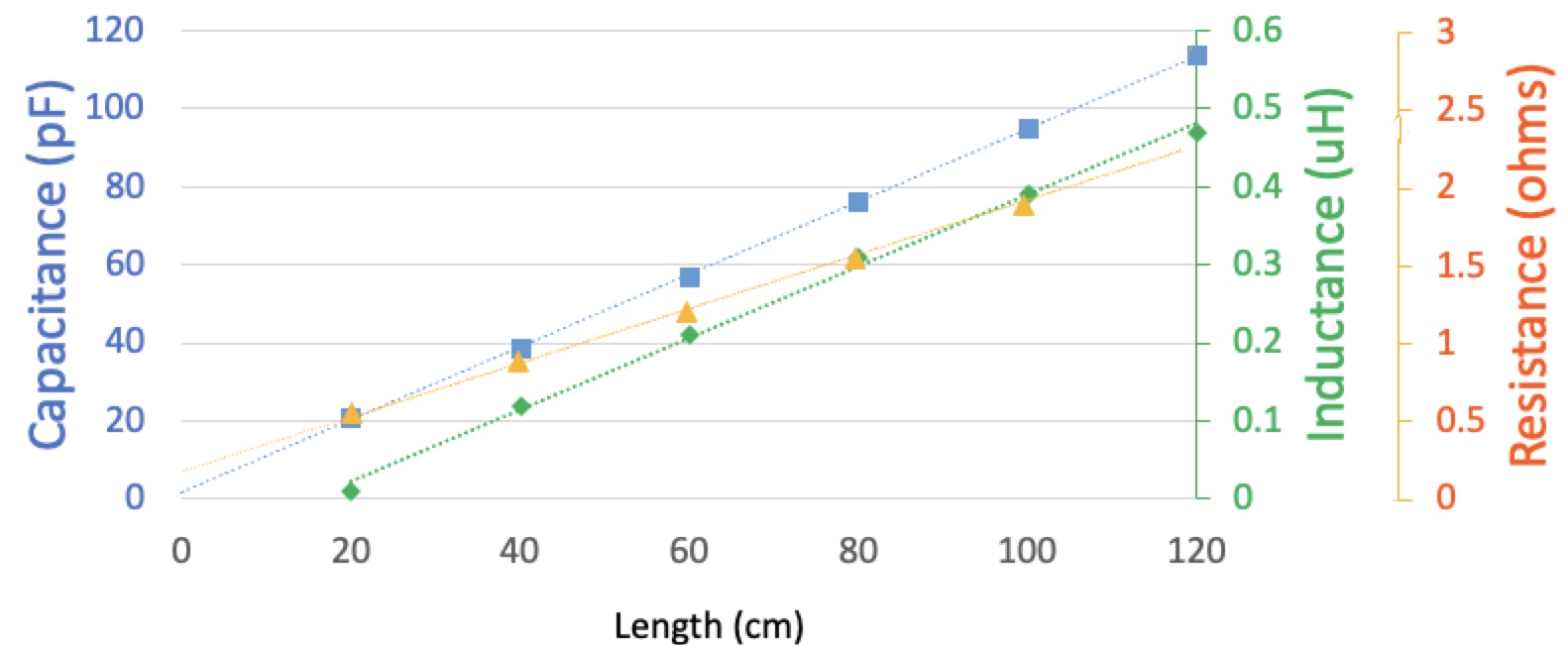}
    \caption{The capacitance, inductance and resistance of the twisted transmission lines shown by the length.}
    \Description{This figure demonstrates the capacitance, inductance and resistance of twisted transmission lines with length ranging from 0 to 120 cm.}
    \label{fig:fig5}
\end{figure}

To inform the following design and implementation of smart textile interfaces, we additionally conducted an experiment to measure the resistance, capacitance, and inductance of twisted transmission lines ranging from 200mm to 1200mm on the human body, with results shown in Figure \ref{fig:fig5}. The properties showed a strong correlation with length ($R^2 = 0.99$). For the longest line (1200mm), the capacitance was 113.5pF, which could limit impedance changes in the high-frequency range. However, we confirmed that the low-frequency range is sufficient to support up to three sensor circuits, as discussed in Section 5.

\section{Sensor Value Estimation}

Once the coil and transmission line designs for smart textile interfaces were determined, another key challenge was the development of an algorithm to extract sensor values from the measured S11 spectrum.
\MK{
While it is possible to use an optimization algorithm to estimate unknown sensor values by iterating through all possible parameters to find the best fit to the mathematical impedance model, this approach was ineffective for two reasons.
First, although many variables such as $L_t$, $L_r$, $C_{SMA}$, and other preset component values in sensor circuits can be assumed known in the fabrication process, the mathematical model still involved too many unknown parameters, such as $k$, $C_{line}$ and each sensor value. Thus, directly using optimization algorithms often leads to convergence to local minima and is unstable in finding the true global solution.
Second, the mobile impedance reader was typically limited in capturing a high-resolution, high-precision impedance spectrum, which constrained both the quality and quantity of data points available for optimization algorithms. This limitation could cause the fitting process to fail, as the optimization might converge to inaccurate solutions due to insufficient detailed information.
These limitations necessitated alternative approaches to reliably extract sensor values from the measured impedance spectrum.}

 To overcome this challenge, our core idea is to use impedance at the resonant frequency of each sensor circuit to predetermine some factors in steps. 
\MK{This is because at resonance, the sensor circuit's impedance is near zero if the resistance is low. This creates a short circuit in the smart textile interface, simplifying the circuit model and allowing the equations to be approximated as follows:}
\begin{gather}
    \label{eq_rf}
    \lvert Z(f) \rvert = \lvert\frac{1}{\frac{1}{(2\pi fL_t)j-\frac{((2\pi f)k\sqrt{(L_t L_r)}j)^2}{ 2\pi f(L_{line_i} + L_r)j}}+2\pi fC_{SMA}j}\rvert 
\end{gather}
where $i$ indicates the ith sensor circuit in which the resonance occurs and $L_t$, $L_r$, $C_{SMA}$, $L_{line_i}$ and $R_{line_i}$ are known during fabrication process \MK{(see Appendix \ref{sec_appendix_1} for approximation details).}

By leveraging this idea, we additionally incorporated a reference circuit with known LC components into the smart textile interface. Since the values of the LC components in the reference circuit were predetermined, we can directly obtain its resonant frequency and use the resonant frequency as a starting point to navigate the challenges in estimating sensor values.
\MK{As a result, our algorithm can be divided into the following three steps.}

\subsection{Step 1: Calculating Coupling Factor}
In light of Eq \ref{eq_rf}, 
we can calculate the coupling factor directly at the resonant frequency of the reference LC circuit, without needing to estimate sensor values and the capacitance of each transmission line ($C_{line_i}$). However, to do this, we must first obtain the impedance at that resonant frequency. Due to the limited number of data points measured from the reader, the S11 value at the exact resonant frequency may not be directly available. To address this, we used a linear interpolation technique to estimate the S11 value at the desired frequency. The system then converted the S11 value to the impedance at the resonant frequency, allowing the coupling factor to be calculated using Eq. \ref{eq_rf}. 

\subsection{Step 2: Estimating Capacitive or Inductive Sensor Values}
\begin{figure}
    \centering
    \includegraphics[width=1\linewidth]{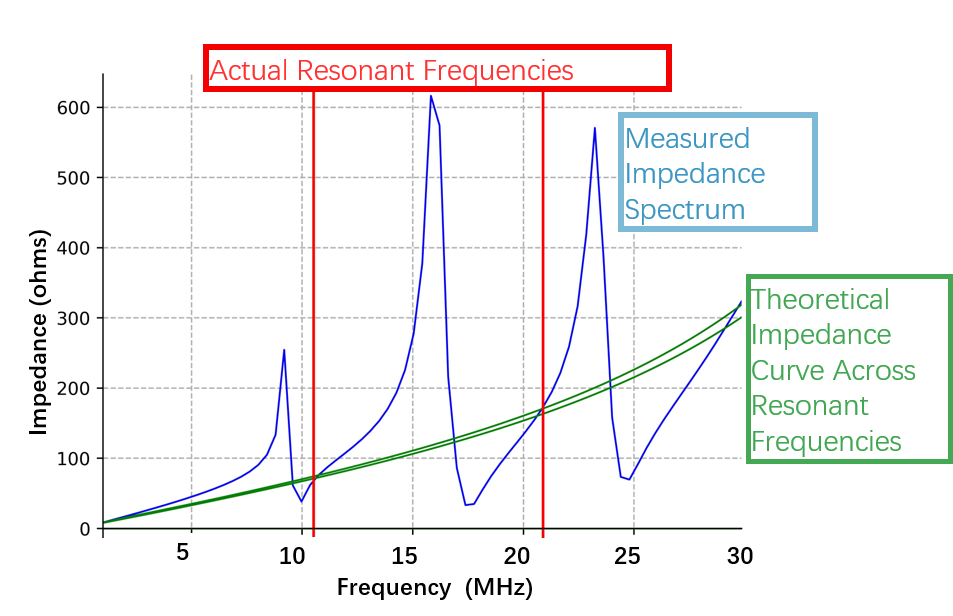}
    \caption{The absolute impedance spectrum of the entire system and the supposed resonant frequencies (red line) of each sensor circuit used in the model validation. \MK{The green lines illustrate the value of Eq. \ref{eq_rf} calculated by the two sensor circuits with lower resonant frequency}. Their intersections with the measured impedance spectrum accurately represent the resonant frequencies of the sensor circuits. \MK{Note that the resonant frequency of the third sensor circuit is 30.9MHz, which exceeds 30MHz and is not plotted.}}
    \Description{
    This figure demonstrates the impedance of the prototype and the relationship between the estimated intersections of plots and the actual resonant frequencies.The absolute impedance spectrum of the entire system and the supposed resonant frequencies (red line) of each sensor circuit used in the model validation. \MK{The green lines illustrate the value of Eq. \ref{eq_rf} calculated by the two sensor circuits with lower resonant frequency}. Their intersections with the measured impedance spectrum accurately represent the resonant frequencies of the sensor circuits. \MK{Note that the resonant frequency of the third sensor circuit is 30.9MHz, which exceeds 30MHz and is not plotted.
    }}
    \label{fig:rf_theoretical_curve}
\end{figure}

Once the coupling factor ($k$) was estimated, our second step was to identify the resonant frequency of each sensor circuit and derive the corresponding capacitive or inductive sensor values. This step was challenging, as the lowest peaks in the absolute impedance spectrum do not correspond to the actual resonant frequencies of these sensor circuits, as illustrated by the red lines in \MK{Figure \ref{fig:rf_theoretical_curve}}. \MK{
The reactive components, including the inductance of the receiver and transmitter coils and the impedance of the transmission lines, interact with sensor circuits significantly, causing complex shifts in impedance spectrum (see Appendix \ref{sec_appendix_1} for details).}
To accurately find the resonant frequency of each RLC circuit, we applied Eq. \ref{eq_rf} in reverse.

Specifically, given the coupling factor ($k$) is known, Eq. \ref{eq_rf} becomes an equation with a single variable, the frequency ($f$). Thus, our algorithm can plot Eq. \ref{eq_rf} across the frequency spectrum for each sensor circuit and search for intersections with the impedance spectrum that was converted and interpolated from the measured S11 spectrum, \MK{(as shown by the green lines in Figure \ref{fig:rf_theoretical_curve})}. 
The frequencies at these intersections are potential resonant frequencies for the sensor circuits, as they satisfy Eq. \ref{eq_rf}. 
To determine which frequency is the resonant frequency of each sensor circuit, the algorithm seeks for the points where the impedance trend increases. This is because this increasing trend indicates that the system is approaching to another resonance due to the sensor circuit. Finally, the algorithm select the frequency closest to the last estimated resonant frequency as the resonant frequency for each sensor circuit.  \MK{However, it is important to note that  for sensor circuits incorporating resistive sensors, we assumed that their resonant frequencies are predetermined during the fabrication process and therefore do not require estimation.}

Once the resonant frequency is determined, our algorithm can easily derive the capacitive or inductive sensor value in each sensor circuit using the formula of the resonant frequency of an RLC circuit \cite{lara2022series}:

\begin{gather}
    f = \frac{1}{2\pi \sqrt{l_ic_i}}
\end{gather}

where one of $l_i$ and $c_i$ is assumed to be known during the fabrication process and the other is the sensor value.

\subsection{Step 3: Approximating Transmission Line Capacitance and Resistive Sensor Values}
After determining the capacitive and inductive sensor values in each sensor circuit, our final step was to estimate resistive sensor values. Unlike the previous steps, there were no alternative methods or shortcuts available to estimate the resistive sensor values. We were still required to solve the capacitance of each transmission line ($C_{line_i}$) first before proceeding with the resistive value approximation. 

Our strategy for estimating $C_{line_i}$ was adjusting each $C_{line_i}$ until the lowest peaks' frequencies from the predicted S11 spectrum align with those from measure S11 spectrum. This strategy was designed due to the insight that the value of $C_{line_i}$ can significantly influence the peak positions in the S11 or impedance spectrum, as these peaks occur when all reactive components in the circuits of the smart textile interface cancel each other out. However, it is important to note that this estimation may be not accurate enough due to the limited resolution of the reader's measurements.  

Once the transmission line capacitance was roughly estimated, the algorithm shifted to approximate resistive sensor values. To achieve this, we employed a regression fit optimization algorithm (i.e. Trust Region Reflective algorithm in our implementation) to search for the best-fitting resistive sensor values that would allow the predicted S11 spectrum to closely align with the measured S11 spectrum. The initial guess for this optimization algorithm was based on the last estimated resistive values.  Simultaneously, the algorithm refined the transmission line capacitance as well, to further improve the alignment between the measured spectrum and the estimated spectrum. This approach allowed the algorithm to approximate the resistive sensor values with reasonable accuracy. Note that this accuracy was primarily dependent on the precision and resolution of S11 or impedance spectrum. Currently, we focused on the S11 spectrum because it was directly measured by our reader. Converting the S11 spectrum to the impedance spectrum may introduce inaccuracies, as high impedance values result in only minor changes in the S11 spectrum, making them difficult to detect with the reader's limited resolution.

\section{Simulation-based Evaluation}
\label{sec_simulation_based_evaluation}


To assess the accuracy of the sensor value extraction algorithm under varying conditions, we conducted a simulation-based evaluation. \MK{This simulation accounted for real-world factors such as device sampling resolution and data noise to mimic actual collected data. This approach was chosen due to the vast number of potential circuit configurations and conditions in a smart textile interface, making physical testing impractical. Simulations enabled us to gain a deeper understanding of the algorithm’s performance across different scenarios (e.g., varying transmission line lengths and sensor values) and provided valuable insights for future system design, optimization, and implementation.}


\subsection{Experiment 1: Coupling Factor Estimation}
The first step in our sensor value estimation algorithm was calculating the coupling factor. This experiment aimed to assess the accuracy of the calculated coupling factor across different reference circuit configurations and transmission line lengths in a smart textile interface, providing insights for designing the reference circuit.

\subsubsection{Method}
\MK{To evaluate the performance of coupling factor estimation, we developed a simulator to generate multiple S11 spectra simulating measurements from the reader on various circuit setups. Using these spectra, we calculated coupling factors with our algorithm and compared them to ground-truth values to assess accuracy.}

\MK{The simulator used Eq. 1 to Eq. 6, validated through Section \ref{sec_model_validation}. To simplify the process, we assumed the smart textile interface only involved the reference circuit with an open-circuited transmission line, as other sensor circuits would be designed to avoid overlap with the reference circuit's spectrum.}

\MK{The parameters $L_{t}$, $C_{SMA}$, and $L_{r}$ were set to 0.6$\mu H$, 10pF, and 4.54$\mu H$ based on coil design studies. The coupling factor ranged from 0.25 to 0.29 (perfect and weak alignment) randomly. Transmission line lengths were varied across four ranges: <25cm, 25-50cm, 50-75cm, and 75-100cm. For each range, the simulator randomly selected a length and generated the corresponding $C_{line}$, $L_{line}$, and $R_{line}$ values according to the transmission line design results. We simulated resonant frequencies of the reference circuit from 1 MHz to 30 MHz in 100 steps, fixing the inductance-to-capacitance ratio at 1, and calculated the specific capacitance and inductance values.}

\MK{To simulate real-world conditions, we limited the sampling resolution to 101 points across the spectrum and added Gaussian noise with three decimal places to each S11 value, reflecting the reader's resolution and precision. Each condition was repeated 1,000 times to account for randomness. In total, the simulator generated 4 transmission line length ranges × 100 resonant frequencies × 1,000 repetitions, producing 4,000,000 S11 spectra for analysis. For each spectrum, we applied the first step of our algorithm to estimate the coupling factor, assessing accuracy by comparing the estimated value to the ground truth.}

\subsubsection{Results}
\MK{We averaged the accuracies across 1,000 repetitions and presented the results in Figure \MK{\ref{fig:cr_sensing_accuracy}a}. We found that transmission line length had no significant impact on accuracy, but the coupling factor estimation was less stable when the reference circuit’s resonant frequency was below 10 MHz, ranging from 84\% to 98\%. This instability occurred because, at lower frequencies,  the S11 values exhibited more pronounced changes when reactive components cancel out each other. If the sampling missed these changes, the interpolated spectrum became less accurate, leading to discrepancies in the coupling factor. Above 10 MHz, accuracy stabilized at about 99\%. We recommend setting the reference circuit’s resonant frequency above 10 MHz for optimal accuracy and stability.}



\subsection{Experiment 2: Sensor Value Estimation for Single Sensor}
Next, we validated the accuracy of sensor value estimation with a single sensor integrated into the smart textile interface. We used a similar simulation approach to test accuracy under different sensor circuit configurations and varying transmission line lengths. Coil alignments were not tested, as previous experiments already demonstrated high accuracy in coupling factor estimation.


\subsubsection{Method}
\MK{We modified the simulator from the previous experiment to meet the objectives of this one. First, we standardized the reference circuit design with a resonant frequency of 27 MHz and an inductance-to-capacitance ratio of 1. According to results from previous experiments, this configuration can provide accurate estimates of the coupling factor. Second, while the transmission line length was still randomly selected within the four defined ranges, we introduced ±20\% fluctuations in transmission line capacitance to simulate real-world conditions such as bending and folding. Our algorithm accounted for the initial capacitance, as the transmission line length is known during fabrication. We kept the resistance and inductance of the transmission line stable, as previous studies showed these parameters did not vary significantly under different conditions.
We also added a sensor circuit, varying its resonant frequency from 1 MHz to 25 MHz in 100 steps and randomly assigning the inductance-to-capacitance ratio between 0.1 and 2. This setup allowed simulation of various sensor values. We capped the resonant frequency at 25 MHz to avoid overlap with the reference circuit’s frequency. The sensor circuit’s resistance ranged from 10 ohms to 60 ohms, varying ±50\% in each iteration to simulate changes in resistive sensor values. The other settings, including transmission line ranges and Gaussian noise, were the same as in the previous experiment.}

In total, the simulator generated 4,000,000 S11 spectra (4 ranges × 100 frequencies × 1,000 repetitions). For each spectrum, we first calculated the coupling factor and then estimated the sensor values. Since accuracy for capacitive and inductive sensor values is expected to be identical (Eq. 8), we focused on capacitive sensor value estimation. Accuracy was determined by dividing the estimated sensor value by the set value.


\begin{figure*}[htbp]
    \centering
    \includegraphics[width=1.0\linewidth]{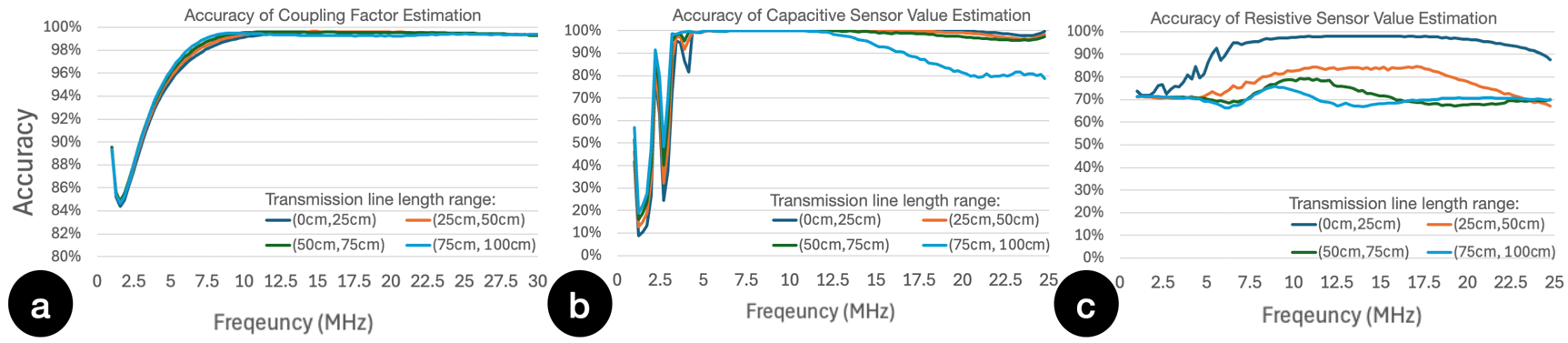}
    \caption{(a) The accuracy of coupling factor estimation across frequencies ranging from 1 MHz to 30 MHz (b) The accuracy of estimating single capacitive sensor value across frequencies ranging from 1 MHz to 25 MHz, considering four different ranges of transmission line lengths. (c) The accuracy of estimating single resistive sensor value across frequencies ranging from 1 MHz to 25 MHz, considering four different ranges of transmission line lengths.  }
    \Description{
    This figure demonstrates the result from experiments in section 5.
    (a) The accuracy of coupling factor estimation across frequencies ranging from 1 MHz to 30 MHz (b) The accuracy of estimating single capacitive sensor value across frequencies ranging from 1 MHz to 25 MHz, considering four different ranges of transmission line lengths. (c) The accuracy of estimating single resistive sensor value across frequencies ranging from 1 MHz to 25 MHz, considering four different ranges of transmission line lengths.  
    }
    \label{fig:cr_sensing_accuracy}
\end{figure*}

\subsubsection{Results}
\MK{We analyzed the accuracy of capacitive and resistive sensor value estimations separately, as they were addressed in different algorithm steps.}

\MK{Figure \ref{fig:cr_sensing_accuracy}b shows the accuracy of capacitive sensor value estimation. Below 5 MHz, accuracy was unstable, similar to the coupling factor estimation. Above 5 MHz, accuracy stabilized at 99\%. However, longer transmission lines reduced accuracy, particularly above 25 MHz, as longer lines increased capacitance, causing effects like short circuits. For transmission lines over 50 cm, accuracy dropped below 90\% at high frequencies, emphasizing the importance of accounting for transmission line length when configuring sensor circuits.}

\MK{For resistive sensor value estimation, Figure \ref{fig:cr_sensing_accuracy}c shows similar trends. Accuracy was low below 5 MHz but improved above 5 MHz. When transmission lines were under 25 cm, accuracy exceeded 90\% from 5 MHz to 24 MHz, but dropped with longer lines. Experimenting with inductance-to-capacitance ratios, we found that limiting the ratio to 0.5 improved accuracy, especially for longer transmission lines, e.g., from 84\% to 93\% in the 50-75 cm range. Lowering the ratio helped mitigate accuracy loss with longer lines, improving resistive sensor estimation performance.}


\MK{Lastly, we calculated transmission line capacitance estimation accuracy, which averaged 97\% across configurations and line lengths. This indicates the algorithm’s effectiveness in estimating capacitance, which could be useful to monitor line conditions such as bending or physical disturbances, enabling potential applications in activity sensing \cite{yu2024seamposerepurposingseamscapacitive}.}

\subsection{Experiment 3: Sensor Value Estimation for Multiple Sensors}
Estimating sensor values in the concurrent operation of multiple sensors posed additional challenges compared to a single sensor. For example, if the resonant frequencies of multiple sensor circuits were too close, interference could complicate estimation. In our experiment, we evaluated the performance of our algorithm with three sensors embedded in the smart textile interface, using a similar simulation-based approach to assess the impact of different circuit configurations on the accuracy of multi-sensor value estimation.

\subsubsection{Method}
\MK{We used the same simulator as in the previous experiment, incorporating three sensor circuits into the smart textile interface. To assess how different configurations affect accuracy, we varied the resonant frequency of one sensor circuit from 5 MHz to 20 MHz in 100 steps, adjusting the gap between its resonant frequency and those of the other two circuits. The frequency gap ranged from 1 MHz to 5 MHz in 1 MHz increments, with all sensor circuits constrained to a range between 1 MHz and 25 MHz. The inductance-to-capacitance ratio of each sensor circuit was randomly assigned between 0.1 and 2.
We limited the transmission line length to 0-25 cm, as longer lines were shown to reduce the available spectrum in the previous experiment. The modified simulator generated 500,000 S11 spectra for analysis (100 frequencies × 5 gaps × 1,000 repetitions).
We focused on estimating the sensor values of the circuit with the middle resonant frequency, as it was most affected by the neighboring circuits. The same algorithm was applied to estimate sensor values, allowing comparison with the single-sensor scenarios.}

\begin{figure*}[htbp]
    \centering
    \includegraphics[width=1.0\linewidth]{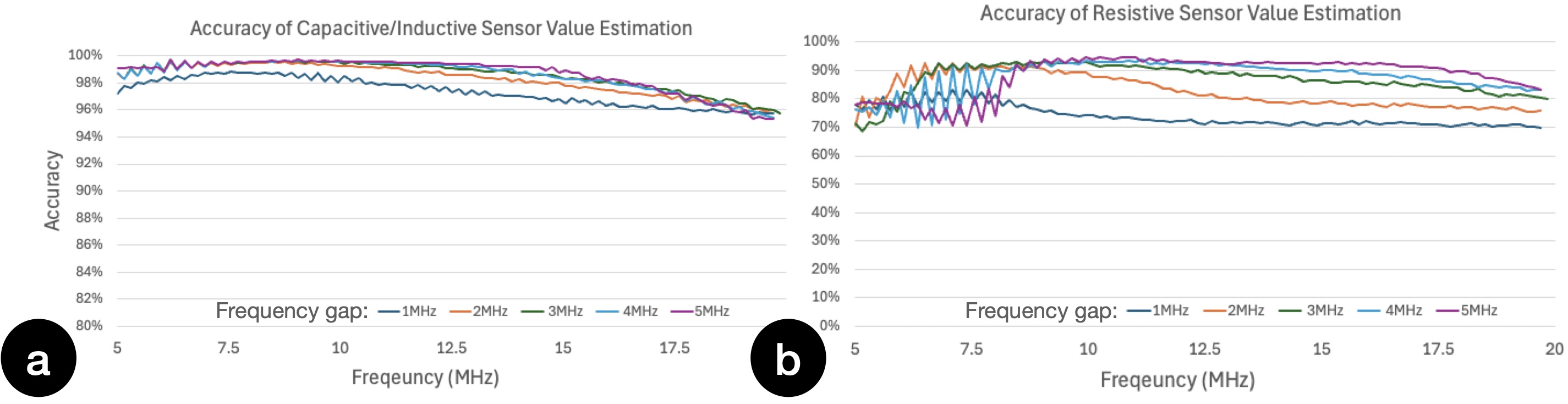}
    \caption{The results of sensor value estimation when three sensor circuits were integrated into a smart textile interface. (a) The accuracy of estimating capacitive/inductive sensor values across frequencies ranging from 5 MHz to 20 MHz, considering different frequency gap. (b) The accuracy of estimating resistive sensor values across frequencies ranging from 5 MHz to 20 MHz, considering different frequency gap.  }
    \Description{
    This figure demonstrates the result from experiments in section 5.
    The results of sensor value estimation when three sensor circuits were integrated into a smart textile interface. (a) The accuracy of estimating capacitive/inductive sensor values across frequencies ranging from 5 MHz to 20 MHz, considering different frequency gap. (b) The accuracy of estimating resistive sensor values across frequencies ranging from 5 MHz to 20 MHz, considering different frequency gap.}
\label{fig:multiple_c_r_sensing_accuracy}
\end{figure*}
\subsubsection{Results}

\MK{
Figure \ref{fig:multiple_c_r_sensing_accuracy}a shows the accuracy of capacitive sensor value estimation. As expected, smaller frequency gaps slightly reduced accuracy. For example, with a 1 MHz gap, accuracy averaged 97\%, while a 5 MHz gap resulted in 98\%, closely matching the single-sensor scenario at 99\%. The middle sensor circuit, influenced by both neighboring circuits, was the most challenging, but the other two circuits, with less interference, showed better accuracy. These results suggest the potential for expanding to more than three sensors in a smart textile interface. For instance, with a 2 MHz gap, up to 12 sensors could fit within the 1 MHz to 25 MHz range, though further validation is needed.}

\MK{Figure \ref{fig:multiple_c_r_sensing_accuracy}b shows resistive sensor value estimation accuracy. Below 10 MHz, accuracy was unstable due to difficulties estimating resonant frequency and transmission line capacitance, especially when one sensor circuit had a resonant frequency below 5 MHz. Above 10 MHz, accuracy stabilized. With a 1 MHz frequency gap, accuracy dropped to 74\%, indicating increased interference. However, with a 5 MHz gap, accuracy rose to 88\%, which is close to the 92\% accuracy of the single-sensor scenario. This shows that our system can maintain high accuracy with multiple sensors if the frequency gap is sufficiently large.}

\subsection{Experiment 4: Comparison with Real-World Performance}
\MK{To understand how the simulation study results deviate from real-world performance, we implemented a physical prototype and compared its accuracy in estimating the coupling factor ($k$), sensor capacitance ($c$), and resistance ($r$) with the simulated test results.}

\subsubsection{Apparatus}
\MK{We created a smart textile interface prototype using the selected coil and transmission line designs. The prototype included three sensor circuits with resonant frequencies of 9.9 MHz, 14.5 MHz, and 19.9 MHz, each connected to a 30 cm transmission line.
To simulate changes in capacitive or resistive sensor values, we prepared a set of fixed capacitors and resistors, enabling controlled adjustments to the resonant frequency of the middle-frequency sensor circuit from 12.2 MHz to 16.6 MHz and resistance from 10 to 50 ohms in five steps. We implemented a reader using NanoVNA and the chosen transmitter coil design. Aligning the reader's coil randomly with the interface's coil within 10mm, we measured the S11 spectrum from 5 MHz to 30 MHz with 101 data points. This setup was designed to mirror the simulation environment, ensuring consistency between simulated and physical test conditions.}

\subsubsection{Method}
\MK{To account for human body influence, we recruited 10 participants (9 male, 1 female) and attached the prototype to their backs, similar to Figure 3. We then randomly altered the coil alignment five times per participant, measuring the coupling factor using the standard approach \cite{Jeon2019} for each alignment. In addition, we adjusted the capacitor and resistor in the middle-frequency sensor circuit to have five levels of capacitance and resistance. For each configuration, we collected 20 S11 spectra to account for possible impacts caused by body postures. In total, we collected 3000 impedance spectra (10 people $\times$ (5 coupling factor + 5 capacitance + 5 resistance) $\times$ 20 repetitions) to evaluate the accuracy of estimating the coupling factor, capacitance, and resistance.}

\subsubsection{Results}
\MK{
The average accuracies for estimating the coupling factor ($k$) and capacitance ($c$) were \MG{98\% and 96\%}, similar to the simulation results of 99\% and 98\% respectively. \MG{When converted to actual capacitance, the mean absolute error (MAE) of capacitance estimation was 0.45$pF$}. For resistance estimation, the average accuracy across all participants was 91\%, ranging from 87\% to 94\%, with a mean absolute error of 2.2 ohms. These results aligned with the simulation outcomes (88\% for the 5 MHz gap and 92\% for single sensor estimation). 
\MG{The estimation of resistance exhibited a lower degree of accuracy in contrast to the capacitance estimation, potentially attributable to inaccuracies in the fitting process and the reader’s limited resolution.}
\MG{Nevertheless, the accuracy levels for both resistance and capacitance estimation were sufficient for common textile sensors \cite{Parzer2018, Aigner2021} to achieve activity detection, as the changes in sensor values exceeded the estimation error. This was further validated through our user study in Section \ref{sec_evaluation}}.
Overall, \MG{the result of the real-world experiments} indicates that the simulation results closely matched the real-world performance of this prototype, validating the reliability of our model and the insights gained from the simulation study.}

\section{IMPLEMENTATION}

In this section, we will discuss the detailed implementation of our smart textile interface and operating system.

\subsection{Battery-free, IC-less and Wireless Smart Textile Interface}
\label{sec_implementation_of_smart_textile_interface}
\MK{Our smart textile interface, consisting of a receiver coil, transmission lines, constant RLC components, and textile sensors, was implemented on a Muslin Fabric Cotton substrate \cite{MuslinFabric2020}. We opted 34AWG Litz Wire \cite{BNTECHGO2022}, a common choice for low-resistance and insulated wiring \cite{10.1145/3613904.3642528, 10.1145/3294109.3295625, 10.1145/3613905.3648646} to fabricate the receiver coil and transmission lines. We employed a Brother SE600 embroidery machine \cite{BrotherSE600} to embroider the wire as bobbin thread through straight stitches with the stitch length of 2.5mm. For constant RLC components, we used 2-pin 0604 SMD components soldered onto the Litz wires, following the method described in \cite{Molla2017}. Since SMD inductors typically have a low Q factor, we replaced them with embroidered or I-shaped coil inductors when higher inductance was needed.}

\begin{figure*}
    \centering
    \includegraphics[width=1.0\linewidth]{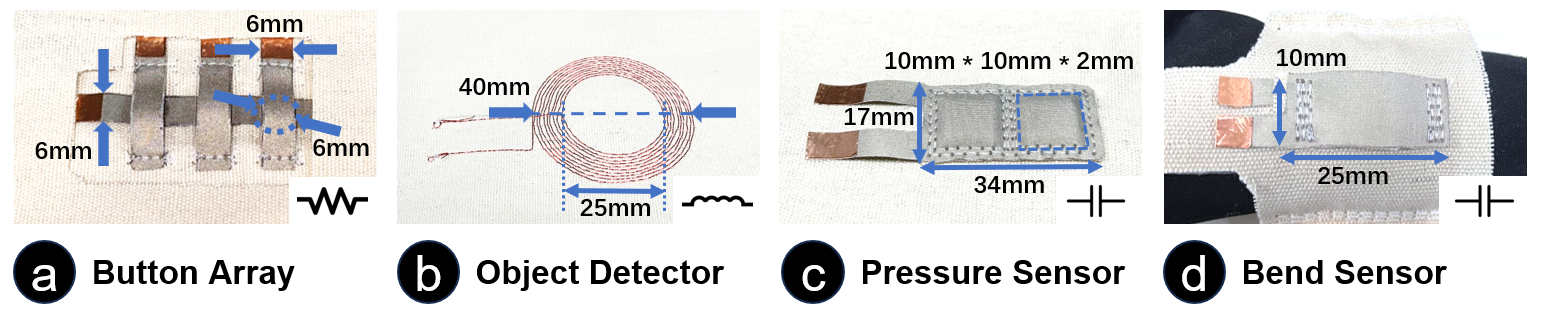}
    \caption{
    Four representative textile sensors implemented in our battery-free and IC-less smart textile interface.
    \MK{
    \textbf{(a)} A resistive button array consisting of 3 pressure-sensitive buttons. Each button is constructed from two pieces of conductive fabric strip (width = 6mm)\cite{ConductiveFabric} sandwiching a circular piece of resistive foam (thickness = 1mm)\cite{ResistiveFoam}. The resistance of the button array is 60$\Omega$ with no pressure. When each button is pressed, the resistance will drop to 0$\Omega$ to 15$\Omega$, 15$\Omega$ to 30$\Omega$ and 30$\Omega$ to 45$\Omega$  respectively according to the pressure applied.
    \textbf{(b)} An inductive object detector consisting of a 10-turns coil (25mm by 40mm) with an inductance of 5.4 uH. The inductance will decrease accordingly when metallic object (e.g. a key, a card embedded with coils) approaches.
    \textbf{(c)} A capacitive pressure sensor constructed by two pieces of conductive fabric (17mm by 34mm) sandwiching two 10mm × 10 mm × 2mm square pieces of resistive rubber \cite{ConductiveRubber}. The two conductive layers are isolated by a layer of cotton fabric \cite{MuslinFabric2020}. The default capacitance is 15pF and will increase to 20 pF when 2kg pressure is applied.
    \textbf{(d)} a capacitive bend sensor constructed by two pieces of conductive fabric (10mm by 25mm) sandwiching a piece of resistive foam (thickness = 1mm). The two conductive layers are isolated by a layer of cotton fabric. When the sensor is bent from 0 to 90 degrees, the capacitance will increase from 7pF to 15pF.
    }}

    \Description{
    Our battery-free and IC-less smart textile interface supports various types of sensors, including resistive, capacitive and inductive sensors. To demonstrate its capability, we designed and implemented four types of sensors.
    \MK{
    \textbf{(a)} A resistive button array consisting of 3 pressure-sensitive buttons. Each button is constructed from two pieces of conductive fabric strip (width = 6mm)\cite{ConductiveFabric} sandwiching a circular piece of resistive foam (thickness = 1mm)\cite{ResistiveFoam}. The resistance of the button array is 60$\Omega$ with no pressure. When each button is pressed, the resistance will drop to 0$\Omega$ to 15$\Omega$, 15$\Omega$ to 30$\Omega$ and 30$\Omega$ to 45$\Omega$  respectively according to the pressure applied.
    \textbf{(b)} An inductive object detector consisting of a 10-turns coil (25mm by 40mm) with an inductance of 5.4 uH. The inductance will decrease accordingly when metallic object (e.g. a key, a card embedded with coils) approaches.
    \textbf{(c)} A capacitive pressure sensor constructed by two pieces of conductive fabric (17mm by 34mm) sandwiching two 10mm × 10 mm × 2mm square pieces of resistive rubber \cite{ConductiveRubber}. The two conductive layers are isolated by a layer of cotton fabric \cite{MuslinFabric2020}. The default capacitance is 15pF and will increase to 20 pF when 2kg pressure is applied.
    \textbf{(d)} a capacitive bend sensor constructed by two pieces of conductive fabric (10mm by 25mm) sandwiching a piece of resistive foam (thickness = 1mm). The two conductive layers are isolated by a layer of cotton fabric. When the sensor is bent from 0 to 90 degrees, the capacitance will increase from 7pF to 15pF.
    }}
    \label{fig:sensors}
\end{figure*}

\MK{Our interface supports three types of sensors, including resistive, inductive, and capacitive sensors. To demonstrate its capabilities, we implemented four representative sensors. Their dimensions and fabrication details are shown in Figure \ref{fig:sensors}. Note that typical capacitive sensors operate on either self-capacitance or mutual capacitance principles \cite{Grosse-Puppendahl2017}. Self-capacitance is the capacitance between an electrode and earth ground, while mutual capacitance occurs between two electrodes. Our system is incompatible with self-capacitance sensors due to the absence of a strong earth ground, resulting in minimal capacitance changes. In contrast, mutual capacitance sensors project capacitance changes effectively onto the resonant circuit, making them compatible with our system. Additionally, resistive sensors are commonly used in textile applications due to their robustness in varied environmental conditions, including wet environments where capacitive sensors may struggle.}

\subsection{Reader and Operating System}
\label{sec_implementation_operating_system}
We implemented the reader as part of our system for capturing sensor data. This reader consisted of a NanoVNA, connected to a transmitter coil measuring 10mm by 10mm, as shown in Figure \ref{fig:teaser}. \MK{Based on results of usable frequency range from section \ref{sec_simulation_based_evaluation}, the reader is designed to perform sweeps of frequencies ranging from 5 MHz to 30 MHz with a total of 101 sampling points}. This entire operation is completed within 0.1 seconds, resulting in a sampling rate of 10 Hz for our system in its current state. While this sampling rate may not be considered high, it is sufficient for the requirements of a real-time interactive system. We believe that higher sampling rate can be achieved in future iterations of the device by integrating dedicated frequency modulation chips and optimizing the signal processing pipeline. Once the reader captured the impedance data, it passes them to a laptop (Thinkpad Carbon X1) to process the signals. Our sensor signal extraction process is implemented in Python. We employed the Trust Region Reflective least squares algorithm as our regression fit algorithm throughout the process.



\subsection{Example Applications}

We present two usage scenarios to exemplify the capability of our system. Our demo applications were designed around everyday objects that frequently come into direct contact with mobile or wearable devices, including pockets and gloves. We aim to showcase how our approach can enable these everyday objects to become interactive, while still maintaining their passive nature and operating without the need for external hardware and batteries embedded in textile.

\subsubsection{Shirt}
 We incorporated an inductive object detector, a capacitive pressure sensor, and a resistive button array into a battery-less and IC-less smart textile interface on a shirt (Figure \ref{fig:teaser}a). The inductive object sensor, in conjunction with an capacitor of \MK{9.9} $pF$, operated within the frequency range from \MK{20M} to \MK{25} MHz. This sensor was strategically placed on a \MK{lower} pocket of the shirt to detect metallic objects like keys. On the right shoulder of the shirt, we integrated a capacitive pressure sensor that worked with a inductor of \MK{6.5} $\mu H$. This sensor resonated within the frequency range from \MK{10}M to \MK{15} MHz and was designed to capture the pressure applied to the shoulder region, commonly caused by objects such as a shoulder bag. Additionally, we included a button array sensor on the shirt sleeve. This sensor, in conjunction with an inductor of \MK{23.9} $\mu H$ and an capacitor of \MK{19.1} $p F$, operated at the frequency around \MK{7} MHz. The receiver coil was placed near the front pocket of the shirt, ready for the coupling from the reader.

\MK{This shirt has several applications, including health tracking, where a smartphone app monitors shoulder pressure and reminds the user to relieve it, preventing strain or pain. It can also detect metallic objects, like keys or access cards, in the pocket and trigger notifications if they're left behind. Additionally, the shirt features shortcut buttons on the sleeve for controlling music playback or interacting with an AI assistant.}


\begin{figure}
    \centering
    \includegraphics[width=1\linewidth]{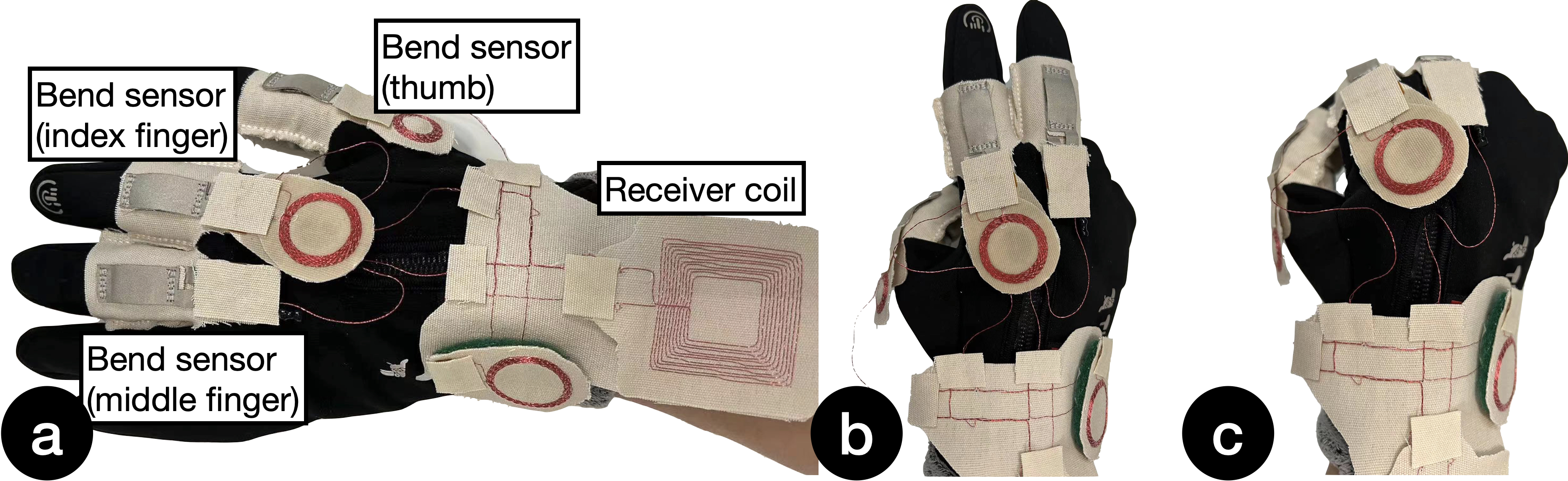}
    \caption{Our glove prototype integrated with BIT. (a) The glove features three bend sensors on the thumb, index, and middle fingers to capture finger gestures. In our implementation, for example, the glove can detect (b) peace and (c) fist gestures.}
    \Description{
    This figure demonstrates the glove prototype integrated with BIT. (a) The glove features three bend sensors on the thumb, index, and middle fingers to capture finger gestures. In our implementation, for example, the glove can detect (b) peace and (c) fist gestures by tracking the bending angle on thumb, index and middle finger.
    }
    \label{fig:glove}
\end{figure}

\subsubsection{Glove}
Although gloves are not typically used for storing personal devices, they share proximity with smartwatches on the user’s wrist. With this in mind, we integrated the smart textile interface into a glove, placing the receiver coil in a position that corresponds to where a smartwatch would typically be situated on the wrist (Figure \ref{fig:glove}a). We incorporated three capacitive bending sensors on the glove to capture the gestures of thumb, index and middle fingers. Each bending sensor has been carefully paired with capacitors and inductors that possess the appropriate values, which allows each sensor to operate within designated frequency ranges.

\MK{This glove can serve as an extension of the input device on the user’s smartwatch, \MK{allowing users to control functions like answering calls with simple gestures such as peace and fist (Figure \ref{fig:glove}b and \ref{fig:glove}c).} Furthermore, the integration of the glove can enhance VR and AR experiences by empowering users to engage in immersive interactions without having to worry about charging the glove.}

\section{User EVALUATION}
\label{sec_evaluation}

We conducted an experiment to assess the effectiveness of our implementation of the interactive shirt prototype.
The primary objective was to measure the accuracy of our approach in detecting various user inputs supported by this prototype. 

\subsection{Participants}

\MK{10 participants were recruited for the study with a mix of 8 males and 2 females, and an average age of 23. All participants are right-handed to facilitate the operation of the sensor placed on the left arm using the dominant hand.} 



\subsection{Procedure} Prior to the study, participants were provided with a concise overview of our prototype. Note that our participants had varying body shapes and typically wore clothing sizes ranging from small to large. During the study, they were instructed to wear our shirt and carry the hardware in the shirt's front pocket. Throughout the study, participants were given the freedom to adopt any posture they deemed comfortable. To evaluate the concurrent operation of the three sensors, we asked participants to perform 60 tasks, with each task simultaneously testing all the three sensors. Specifically, each task consisted of: 1) carrying a shoulder bag to test the pressure sensor, 2) placing an object into a lower pocket to test the object detector, and 3) pressing a button on a sleeve to test the button array. For the bag carrying activity, participants carried either a 0.7 kg or 1.6 kg bag. For the object placement activity, participants placed either an apartment key or a plastic credit card into a pocket. For the button pressing activity, participants pressed one of the three buttons on the sleeve. This resulted in 12 unique combinations (2 bag weights × 2 objects × 3 buttons). Each combination was repeated 5 times, leading to a total of 60 trials. we recorded S11 spectra across the 5 MHz to 30 MHz range using our hardware. Additionally, we collected S11 spectra 10 times while participants were in an idle state to serve as baseline data. In total, 700 S11 spectra were retrieved for analysis, with each participant contributing 10 spectra during the idle state and 60 spectra during the active tasks (10 participants × 70 spectra each).

\begin{figure}
    \centering
    \includegraphics[width=1\linewidth]{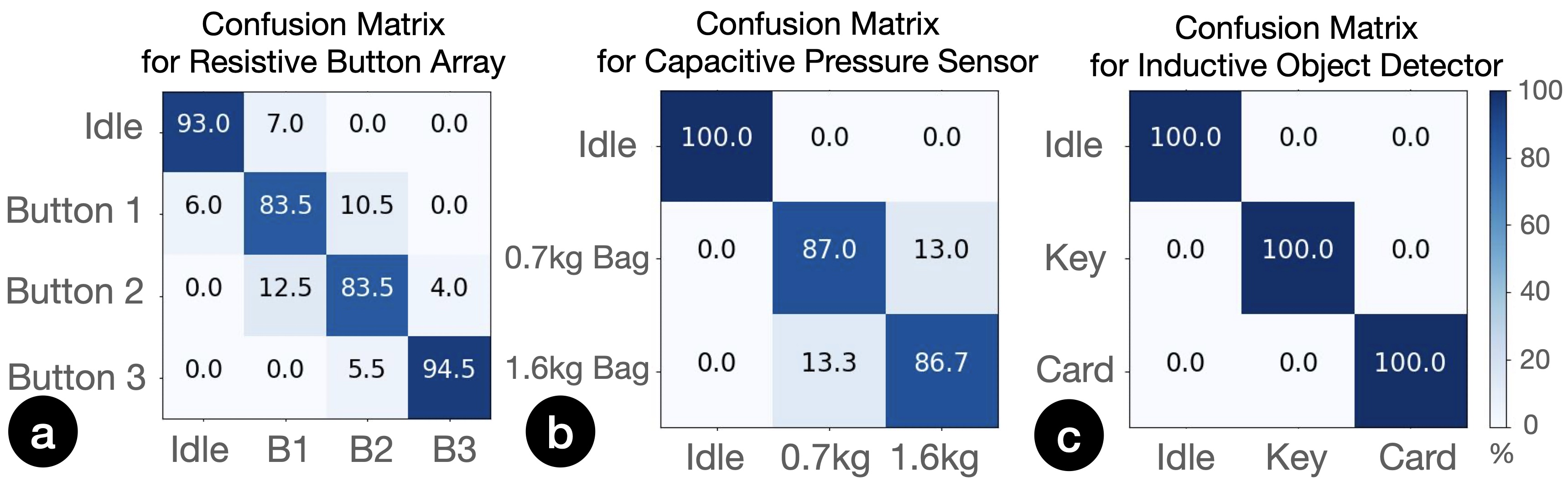}
    \caption{Confusion Matrix for (a) resistive button array, (b) capacitive pressure sensor, (c) inductive object detector.}
    \Description{
    This figure demonstrates the Confusion Matrix for (a) resistive button array, (b) capacitive pressure sensor, (c) inductive object detector.
    }
    \label{fig:study_shirt_results}
\end{figure}

\subsection{Results} 
We employed our algorithm to estimate sensor values for each S11 spectrum. Then, we applied a decision tree classifier to categorize the estimated sensor signals. Due to the considerable variability in body characteristics among participants, the sensor values showed significant variation. Consequently, our analysis emphasized within-subject accuracy, using five-fold cross-validation to validate the classification performance. It is important to note that the variation in sensor signals stems from the sensor itself, which are not our focus. Our objective is to study whether the sensor values estimated from our system are accurate enough for classifying user interactions.

The results are shown in the confusion matrix in Figure \ref{fig:study_shirt_results}.
For the resistive button array, the average accuracy was \MK{88\%}. The confusion matrix indicated significant overlap between the first and second buttons, likely due to participants applying varying levels of force across different trials. This variation may cause their resistance values to become similar. But, if we group these two buttons into the same category, the average accuracy can increase to \MK{93\%}.

On the other hand, the capacitive pressure sensor demonstrated a higher average accuracy of 91\%. The primary source of confusion was from the two tasks of carrying a shoulder bag with two different weights. This is likely because even the same participant might carry the bag differently, causing variations in pressure and sensor readings. Despite that, our estimated capacitive sensor values can still show acceptable accuracy for detecting this interaction \MG{with capacitance changes ranging from 3.0 $pF$ to 6.0 $pF$ across all the cases of carrying the bag (around 1.5 MHz of frequency change between cases)}.

Lastly, the inductive object detector achieved \MK{100\%} accuracy. It demonstrated that the estimated inductive sensor values can reliably distinguish between two different objects and idle state \MG{with inductance changes of around 0.8 $uH$ and 1.5 $uH$ (around 2.5 MHz of frequency change between cases)}. 
\MG{Since the task was less influenced by participant behavior and the sensor robustly established a higher frequency change, the accuracy reached higher value compared with capacitance pressure sensor. This suggests that our system can reliably monitor user interactions given robust and well-designed sensors, without the incorporation of batteries and ICs into textiles.}

In addition, we also examined whether different combinations of tasks resulted in varying accuracy. As a result, no significant differences were observed.

\section{LIMITATIONS AND FUTURE WORK}

In this section, we discuss the limitations of our work and propose potential directions that could further advance it. 

\textit{Data accuracy, readout speed and power consumption of reader.} Our current implementation is limited in terms of accuracy and readout speed, \MK{making it less suitable for applications that require high-speed tracking of subtle sensor changes, such as strain sensors for detailed finger tracking.} This limitation is primarily due to the measurement resolution and sampling rate of NanoVNA. However, we believe that developing a custom device integrated with dedicated frequency modulation chips \cite{Sun2022} and a finely tuned measurement circuit could significantly mitigate this issue. One potential strategy to optimize performance using the custom device is to use focused scanning of pertinent frequency ranges. For instance, we could concentrate impedance measurements solely on relevant resonant frequencies and update the entire spectrum at longer intervals. This approach would allow for enhanced resolution and quicker readings within the sensor's operational frequency range. Another intriguing solution worth exploring is the multitone technique, where multiple frequencies are merged into a singular signal for concurrent impedance analysis. While this method promises faster readout speeds, there may be a potential trade-off in terms of measurement accuracy. Therefore, further investigation is necessary to determine the suitability of this technique, particularly for sensors that prioritize speed over accuracy.

Additionally, power consumption is another critical consideration in the design of the reader. Currently, we used NanoVNA as the reader, which consumes 600mW according to the datasheet \cite{NanoVNA}. This power consumption level could be too high for battery-powered or portable devices. To address this issue, we plan to reduce power consumption. This could involve selecting low-power components, optimizing the measurement circuit \cite{SACRISTANRIQUELME2009177}, and employing power-efficient operating modes. 

\textit{Voltage sensor.} Our approach is based on impedance modulation, which restricts its compatibility with sensors based on voltage, such as microphones, EMG electrodes, and photodiodes. Nevertheless, a potential solution to address this issue is to incorporate a varicap, a compact two-pin component capable of converting voltage signals into variable capacitance. With varicaps, we could enhance our approach to support a wider range of sensors. However, incorporating varicaps presents a significant challenge in achieving high precision and speed in sensor readings. Therefore, our future research will focus on investigating this issue and finding ways to overcome it.


\textit{Cross-textile interface.} Our research is also based on the principle of relay resonators, which have the potential to transfer power across textile surfaces using proper transmitting and receiver coils. However, the challenge is that unlike in our current research, where we could assume that the coils would be well aligned inside a pocket, the coils on the open surfaces of a textile object are subject to movement and instability, making alignment challenging. In order to address this issue, it is essential to investigate relay coil designs that can facilitate alignment, such as using larger or wider coils. This will be a crucial aspect of our future research in this direction.

\textit{Toolkit.} Our method stands out for being battery-free, devoid of integrated circuits, and wireless to operate, which reduces the cost and maintenance of smart textiles. Yet, for the broader adoption of our method, there's a pressing requirement for tools that could facilitate the design and deployment of our solution on textile objects. One promising avenue for future research is a user-friendly software tool specifically designed to assist in designing and implementing receiver coils, sensors, and transmission lines based on individual user interaction demands. For instance, upon inputting desired sensor locations and types, a system like this could automatically generate an optimized design for coils, sensors, and transmission lines. Furthermore, this software tool could also enable the direct conversion of the optimized design into an embroidery file, allowing for the quick realization of the desired idea.

\textit{Pure-textile interface.} Our current implementation requires the inclusion of small, rigid components such as resistors, capacitors, and inductors within textiles. However, with advancements in material science research, it becomes possible to fabricate these rigid components entirely from textile-based materials. Consequently, future textile interfaces could potentially eliminate the need for any rigid components, relying solely on soft materials. This development would signify a significant breakthrough in the field of smart textiles, expanding their range of applications across various use scenarios.

\textit{Hardware integration.} Our current system implementation relies on obtaining the impedance spectrum by measuring the S11 values using a NanoVNA. However, we anticipate that our system can be easily integrated into smartphones and smartwatches in the near future, given that most personal computing devices already come equipped with built-in coils for wireless charging and NFC capabilities. This integration would enable the widespread adoption of smart textiles into everyday life.


\textit{Fabrication with Other Conductive Threads and Fabric Substrates.} \MK{In our implementation, we used Muslin Fabric Cotton as the fabric substrate and 34AWG Litz Wire as the conductive wire for prototyping. However, other textile materials and conductive threads, such as polyester fabrics and silver-coated nylon, may also be used for smart textile interfaces. While we expect that our system should work with these materials, their impact on interface impedance is unknown. For example, different conductive threads may introduce varying resistance and capacitance. Similarly, the fabric substrate can also influence the inductance and capacitance. Therefore, further study is needed to assess how these materials affect system performance in the future.}

\section{CONCLUSION}

This work addresses the challenges associated with incorporating rigid hardware components, such as integrated circuits (ICs), batteries, and connectors, into textile sensors. By leveraging near-field electromagnetic coupling, BIT enables wireless power transfer and data acquisition from textile sensors without the need for traditional hardware embedded in textile. This approach improves usability, reduces manufacturing complexity, and minimizes the environmental impact of textile interfaces. A crucial aspect of our research is the development of a mathematical model and algorithm that take into consideration several challenges, including the influence of transmission lines and coil misalignment, allowing for accurate estimation of sensor readings. Through simulation-based and user-based experiments, we demonstrated the feasibility and versatility of BIT. This research has the potential to transform the landscape of smart textiles, making them more accessible and seamlessly integrated into people's daily lives. By reducing reliance on rigid hardware components, our approach paves the way for a future where smart textiles are not only comfortable to wear but also environmentally sustainable. This work represents a step forward in the evolution of wearable technology, offering new possibilities for innovative applications.

\begin{acks}
This work is supported by the Tsinghua University Initiative Scientific Research Program, the Undergraduate Education Innovation Grants and Florida State University First Year Assistant Professor (FYAP) Grant.
\end{acks}

\begin{thebibliography}{79}


\ifx \showCODEN    \undefined \def \showCODEN     #1{\unskip}     \fi
\ifx \showDOI      \undefined \def \showDOI       #1{#1}\fi
\ifx \showISBNx    \undefined \def \showISBNx     #1{\unskip}     \fi
\ifx \showISBNxiii \undefined \def \showISBNxiii  #1{\unskip}     \fi
\ifx \showISSN     \undefined \def \showISSN      #1{\unskip}     \fi
\ifx \showLCCN     \undefined \def \showLCCN      #1{\unskip}     \fi
\ifx \shownote     \undefined \def \shownote      #1{#1}          \fi
\ifx \showarticletitle \undefined \def \showarticletitle #1{#1}   \fi
\ifx \showURL      \undefined \def \showURL       {\relax}        \fi
\providecommand\bibfield[2]{#2}
\providecommand\bibinfo[2]{#2}
\providecommand\natexlab[1]{#1}
\providecommand\showeprint[2][]{arXiv:#2}

\bibitem[BNT(2023)]%
        {BNTECHGO2022}
 \bibinfo{year}{2023}\natexlab{}.
\newblock \bibinfo{title}{BNTECHGO 34 AWG Magnet Wire - Enameled Copper Wire}.
\newblock \bibinfo{howpublished}{\url{https://a.co/d/8WXGLc8}}.
\newblock


\bibitem[Con(2023)]%
        {ConductiveRubber}
 \bibinfo{year}{2023}\natexlab{}.
\newblock \bibinfo{title}{Conductive Rubber}.
\newblock
  \bibinfo{howpublished}{\url{https://item.taobao.com/item.htm?_u=d208umq5sq61f9&id=593175778341&spm=a1z09.2.0.0.54f02e8drZJAEL}}.
\newblock


\bibitem[DE5(2023)]%
        {DE5000LCRMeter}
 \bibinfo{year}{2023}\natexlab{}.
\newblock \bibinfo{title}{DE-5000 Handheld LCR Meter}.
\newblock \bibinfo{howpublished}{\url{https://a.co/d/6Y5U01U}}.
\newblock


\bibitem[MFR(2023)]%
        {MFRC-522coil}
 \bibinfo{year}{2023}\natexlab{}.
\newblock \bibinfo{title}{MFRC-522 Reader}.
\newblock \bibinfo{howpublished}{\url{
  http://e.tb.cn/h.gpG0cRza9YY6H4i?tk=2OPl364Smic}}.
\newblock


\bibitem[Mus(2023)]%
        {MuslinFabric2020}
 \bibinfo{year}{2023}\natexlab{}.
\newblock \bibinfo{title}{Muslin Fabric Cotton Cloth}.
\newblock \bibinfo{howpublished}{\url{https://a.co/d/1xYwx9R}}.
\newblock


\bibitem[Con(2024)]%
        {ConductiveFabric}
 \bibinfo{year}{2024}\natexlab{}.
\newblock \bibinfo{title}{Conductive Fabric}.
\newblock
  \bibinfo{howpublished}{\url{https://item.taobao.com/item.htm?id=526286030147}}.
\newblock


\bibitem[Nan(2024)]%
        {NanoVNA}
 \bibinfo{year}{2024}\natexlab{}.
\newblock \bibinfo{title}{NanoVNA}.
\newblock \bibinfo{howpublished}{\url{https://nanorfe.com/nanovna-v2.html}}.
\newblock


\bibitem[Res(2024)]%
        {ResistiveFoam}
 \bibinfo{year}{2024}\natexlab{}.
\newblock \bibinfo{title}{Resistive Foam}.
\newblock
  \bibinfo{howpublished}{\url{https://item.taobao.com/item.htm?id=778700605731}}.
\newblock


\bibitem[Bro(2024)]%
        {BrotherSE600}
 \bibinfo{year}{2024}\natexlab{}.
\newblock \bibinfo{title}{SE600 Computerized Sewing and Embroidery Machine with
  4" x 4" Embroidery Area}.
\newblock
  \bibinfo{howpublished}{\url{https://www.brother-usa.com/products/se600}}.
\newblock


\bibitem[SMA(2024)]%
        {SMAconnector}
 \bibinfo{year}{2024}\natexlab{}.
\newblock \bibinfo{title}{SMA Connector}.
\newblock
  \bibinfo{howpublished}{\url{https://detail.tmall.com/item.htm?id=555269804122&ns=1&skuId=3427497713332}}.
\newblock


\bibitem[Ahsan et~al\mbox{.}(2022)]%
        {Ahsan2022}
\bibfield{author}{\bibinfo{person}{M. Ahsan}, \bibinfo{person}{S.H. Teay},
  \bibinfo{person}{A.S.M. Sayem}, {and} \bibinfo{person}{A. Albarbar}.}
  \bibinfo{year}{2022}\natexlab{}.
\newblock \showarticletitle{Smart Clothing Framework for Health Monitoring
  Applications}.
\newblock \bibinfo{journal}{\emph{Signals}} \bibinfo{volume}{3},
  \bibinfo{number}{1} (\bibinfo{date}{Mar} \bibinfo{year}{2022}),
  \bibinfo{pages}{113--145}.
\newblock
\urldef\tempurl%
\url{https://doi.org/10.3390/signals3010009}
\showDOI{\tempurl}


\bibitem[Aigner et~al\mbox{.}(2024)]%
        {10.1145/3613904.3642528}
\bibfield{author}{\bibinfo{person}{Roland Aigner}, \bibinfo{person}{Mira~Alida
  Haberfellner}, {and} \bibinfo{person}{Michael Haller}.}
  \bibinfo{year}{2024}\natexlab{}.
\newblock \showarticletitle{Loopsense: low-scale, unobtrusive, and minimally
  invasive knitted force sensors for multi-modal input, enabled by selective
  loop-meshing}. In \bibinfo{booktitle}{\emph{Proceedings of the 2024 CHI
  Conference on Human Factors in Computing Systems}} (Honolulu, HI, USA)
  \emph{(\bibinfo{series}{CHI '24})}. \bibinfo{publisher}{Association for
  Computing Machinery}, \bibinfo{address}{New York, NY, USA}, Article
  \bibinfo{articleno}{862}, \bibinfo{numpages}{17}~pages.
\newblock
\showISBNx{9798400703300}
\urldef\tempurl%
\url{https://doi.org/10.1145/3613904.3642528}
\showDOI{\tempurl}


\bibitem[Aigner et~al\mbox{.}(2021)]%
        {Aigner2021}
\bibfield{author}{\bibinfo{person}{Roland Aigner}, \bibinfo{person}{Andreas
  Pointner}, \bibinfo{person}{Thomas Preindl}, \bibinfo{person}{Rainer Danner},
  {and} \bibinfo{person}{Michael Haller}.} \bibinfo{year}{2021}\natexlab{}.
\newblock \showarticletitle{TexYZ: Embroidering Enameled Wires for Three
  Degree-of-Freedom Mutual Capacitive Sensing}. In
  \bibinfo{booktitle}{\emph{Proceedings of the 2021 CHI Conference on Human
  Factors in Computing Systems}} (Yokohama, Japan) \emph{(\bibinfo{series}{CHI
  '21})}. \bibinfo{publisher}{Association for Computing Machinery},
  \bibinfo{address}{New York, NY, USA}, Article \bibinfo{articleno}{499},
  \bibinfo{numpages}{12}~pages.
\newblock
\showISBNx{9781450380966}
\urldef\tempurl%
\url{https://doi.org/10.1145/3411764.3445479}
\showDOI{\tempurl}


\bibitem[Aigner et~al\mbox{.}(2020)]%
        {Aigner2020}
\bibfield{author}{\bibinfo{person}{Roland Aigner}, \bibinfo{person}{Andreas
  Pointner}, \bibinfo{person}{Thomas Preindl}, \bibinfo{person}{Patrick
  Parzer}, {and} \bibinfo{person}{Michael Haller}.}
  \bibinfo{year}{2020}\natexlab{}.
\newblock \showarticletitle{Embroidered Resistive Pressure Sensors: A Novel
  Approach for Textile Interfaces}. In \bibinfo{booktitle}{\emph{Proceedings of
  the 2020 CHI Conference on Human Factors in Computing Systems}} (Honolulu,
  HI, USA) \emph{(\bibinfo{series}{CHI '20})}. \bibinfo{publisher}{Association
  for Computing Machinery}, \bibinfo{address}{New York, NY, USA},
  \bibinfo{pages}{1–13}.
\newblock
\showISBNx{9781450367080}
\urldef\tempurl%
\url{https://doi.org/10.1145/3313831.3376305}
\showDOI{\tempurl}


\bibitem[Berglund et~al\mbox{.}(2015)]%
        {Berglund2015}
\bibfield{author}{\bibinfo{person}{Mary~Ellen Berglund}, \bibinfo{person}{Julia
  Duvall}, \bibinfo{person}{Cory Simon}, {and} \bibinfo{person}{Lucy~E.
  Dunne}.} \bibinfo{year}{2015}\natexlab{}.
\newblock \showarticletitle{Surface-mount component attachment for e-textiles}.
  In \bibinfo{booktitle}{\emph{Proceedings of the 2015 ACM International
  Symposium on Wearable Computers}} (Osaka, Japan) \emph{(\bibinfo{series}{ISWC
  '15})}. \bibinfo{publisher}{Association for Computing Machinery},
  \bibinfo{address}{New York, NY, USA}, \bibinfo{pages}{65–66}.
\newblock
\showISBNx{9781450335782}
\urldef\tempurl%
\url{https://doi.org/10.1145/2802083.2808413}
\showDOI{\tempurl}


\bibitem[Blecha({[n.\,d.]})]%
        {Blecha}
\bibfield{author}{\bibinfo{person}{T. Blecha}.}
  \bibinfo{year}{[n.\,d.]}\natexlab{}.
\newblock \bibinfo{title}{Printed and Embroidered Electronic Passive
  Components}.
\newblock
\newblock


\bibitem[Buechley and Eisenberg(2008)]%
        {Buechley2008}
\bibfield{author}{\bibinfo{person}{L. Buechley} {and} \bibinfo{person}{M.
  Eisenberg}.} \bibinfo{year}{2008}\natexlab{}.
\newblock \showarticletitle{The LilyPad Arduino: Toward Wearable Engineering
  for Everyone}.
\newblock \bibinfo{journal}{\emph{IEEE Pervasive Computing}}
  \bibinfo{volume}{7}, \bibinfo{number}{2} (\bibinfo{date}{Apr}
  \bibinfo{year}{2008}), \bibinfo{pages}{12--15}.
\newblock
\urldef\tempurl%
\url{https://doi.org/10.1109/MPRV.2008.38}
\showDOI{\tempurl}


\bibitem[Buechley and Hill(2010)]%
        {Buechley2010}
\bibfield{author}{\bibinfo{person}{Leah Buechley} {and}
  \bibinfo{person}{Benjamin~Mako Hill}.} \bibinfo{year}{2010}\natexlab{}.
\newblock \showarticletitle{LilyPad in the wild: how hardware's long tail is
  supporting new engineering and design communities}. In
  \bibinfo{booktitle}{\emph{Proceedings of the 8th ACM Conference on Designing
  Interactive Systems}} (Aarhus, Denmark) \emph{(\bibinfo{series}{DIS '10})}.
  \bibinfo{publisher}{Association for Computing Machinery},
  \bibinfo{address}{New York, NY, USA}, \bibinfo{pages}{199–207}.
\newblock
\showISBNx{9781450301039}
\urldef\tempurl%
\url{https://doi.org/10.1145/1858171.1858206}
\showDOI{\tempurl}


\bibitem[Charkhabi et~al\mbox{.}(2021)]%
        {Charkhabi2021}
\bibfield{author}{\bibinfo{person}{S. Charkhabi}, \bibinfo{person}{K.J.
  Jackson}, \bibinfo{person}{A.M. Beierle}, \bibinfo{person}{A.R. Carr},
  \bibinfo{person}{E.M. Zellner}, {and} \bibinfo{person}{N.F. Reuel}.}
  \bibinfo{year}{2021}\natexlab{}.
\newblock \showarticletitle{Monitoring Wound Health through Bandages with
  Passive LC Resonant Sensors}.
\newblock \bibinfo{journal}{\emph{ACS Sensors}} \bibinfo{volume}{6},
  \bibinfo{number}{1} (\bibinfo{date}{Jan} \bibinfo{year}{2021}),
  \bibinfo{pages}{111--122}.
\newblock
\urldef\tempurl%
\url{https://doi.org/10.1021/acssensors.0c01912}
\showDOI{\tempurl}


\bibitem[Chen et~al\mbox{.}(2019)]%
        {Chen2019}
\bibfield{author}{\bibinfo{person}{A. Chen}, \bibinfo{person}{J. Tan},
  \bibinfo{person}{X. Tao}, \bibinfo{person}{P. Henry}, {and}
  \bibinfo{person}{Z. Bai}.} \bibinfo{year}{2019}\natexlab{}.
\newblock \showarticletitle{Challenges in Knitted E-textiles}. In
  \bibinfo{booktitle}{\emph{Artificial Intelligence on Fashion and Textiles}}.
  \bibinfo{pages}{129--135}.
\newblock


\bibitem[Cheng et~al\mbox{.}(2013)]%
        {Cheng2013}
\bibfield{author}{\bibinfo{person}{J. Cheng}, \bibinfo{person}{O. Amft},
  \bibinfo{person}{G. Bahle}, {and} \bibinfo{person}{P. Lukowicz}.}
  \bibinfo{year}{2013}\natexlab{}.
\newblock \showarticletitle{Designing Sensitive Wearable Capacitive Sensors for
  Activity Recognition}.
\newblock \bibinfo{journal}{\emph{IEEE Sensors Journal}} \bibinfo{volume}{13},
  \bibinfo{number}{10} (\bibinfo{date}{Oct} \bibinfo{year}{2013}),
  \bibinfo{pages}{3935--3947}.
\newblock
\urldef\tempurl%
\url{https://doi.org/10.1109/JSEN.2013.2259693}
\showDOI{\tempurl}


\bibitem[Devendorf and Di~Lauro(2019)]%
        {10.1145/3294109.3295625}
\bibfield{author}{\bibinfo{person}{Laura Devendorf} {and} \bibinfo{person}{Chad
  Di~Lauro}.} \bibinfo{year}{2019}\natexlab{}.
\newblock \showarticletitle{Adapting Double Weaving and Yarn Plying Techniques
  for Smart Textiles Applications}. In \bibinfo{booktitle}{\emph{Proceedings of
  the Thirteenth International Conference on Tangible, Embedded, and Embodied
  Interaction}} (Tempe, Arizona, USA) \emph{(\bibinfo{series}{TEI '19})}.
  \bibinfo{publisher}{Association for Computing Machinery},
  \bibinfo{address}{New York, NY, USA}, \bibinfo{pages}{77–85}.
\newblock
\showISBNx{9781450361965}
\urldef\tempurl%
\url{https://doi.org/10.1145/3294109.3295625}
\showDOI{\tempurl}


\bibitem[Galli et~al\mbox{.}(2023)]%
        {Galli2023}
\bibfield{author}{\bibinfo{person}{V. Galli}, \bibinfo{person}{S.K. Sailapu},
  \bibinfo{person}{T.J. Cuthbert}, \bibinfo{person}{C. Ahmadizadeh},
  \bibinfo{person}{B.C. Hannigan}, {and} \bibinfo{person}{C. Menon}.}
  \bibinfo{year}{2023}\natexlab{}.
\newblock \showarticletitle{Passive and Wireless All‐Textile Wearable Sensor
  System}.
\newblock \bibinfo{journal}{\emph{Advanced Science}} \bibinfo{volume}{10},
  \bibinfo{number}{22} (\bibinfo{date}{Aug} \bibinfo{year}{2023}),
  \bibinfo{pages}{2206665}.
\newblock
\urldef\tempurl%
\url{https://doi.org/10.1002/advs.202206665}
\showDOI{\tempurl}


\bibitem[Garnier et~al\mbox{.}(2021)]%
        {Garnier2021}
\bibfield{author}{\bibinfo{person}{B. Garnier}, \bibinfo{person}{P. Mariage},
  \bibinfo{person}{F. Rault}, \bibinfo{person}{C. Cochrane}, {and}
  \bibinfo{person}{V. Koncar}.} \bibinfo{year}{2021}\natexlab{}.
\newblock \showarticletitle{Electronic-components less fully textile multiple
  resonant combiners for body-centric near field communication}.
\newblock \bibinfo{journal}{\emph{Scientific Reports}} \bibinfo{volume}{11},
  \bibinfo{number}{1} (\bibinfo{date}{Jan} \bibinfo{year}{2021}),
  \bibinfo{pages}{2159}.
\newblock
\urldef\tempurl%
\url{https://doi.org/10.1038/s41598-021-81246-z}
\showDOI{\tempurl}


\bibitem[Garnier et~al\mbox{.}(2023)]%
        {Garnier2023}
\bibfield{author}{\bibinfo{person}{B. Garnier}, \bibinfo{person}{P. Mariage},
  \bibinfo{person}{F. Rault}, \bibinfo{person}{C. Cochrane}, {and}
  \bibinfo{person}{V. Koncar}.} \bibinfo{year}{2023}\natexlab{}.
\newblock \showarticletitle{Textile dual-band NFC-A4WP (13.56–6.78 MHz)
  combiner for wireless energy and data transmission for connected clothing}.
\newblock \bibinfo{journal}{\emph{Scientific Reports}} \bibinfo{volume}{13},
  \bibinfo{number}{1} (\bibinfo{date}{Apr} \bibinfo{year}{2023}),
  \bibinfo{pages}{5613}.
\newblock
\urldef\tempurl%
\url{https://doi.org/10.1038/s41598-023-31832-0}
\showDOI{\tempurl}


\bibitem[Glauser et~al\mbox{.}(2019)]%
        {Glauser2019}
\bibfield{author}{\bibinfo{person}{Oliver Glauser}, \bibinfo{person}{Daniele
  Panozzo}, \bibinfo{person}{Otmar Hilliges}, {and} \bibinfo{person}{Olga
  Sorkine-Hornung}.} \bibinfo{year}{2019}\natexlab{}.
\newblock \showarticletitle{Deformation Capture via Soft and Stretchable Sensor
  Arrays}.
\newblock \bibinfo{journal}{\emph{ACM Trans. Graph.}} \bibinfo{volume}{38},
  \bibinfo{number}{2}, Article \bibinfo{articleno}{16} (\bibinfo{date}{March}
  \bibinfo{year}{2019}), \bibinfo{numpages}{16}~pages.
\newblock
\showISSN{0730-0301}
\urldef\tempurl%
\url{https://doi.org/10.1145/3311972}
\showDOI{\tempurl}


\bibitem[Gong et~al\mbox{.}(2019)]%
        {Gong2019}
\bibfield{author}{\bibinfo{person}{Jun Gong}, \bibinfo{person}{Yu Wu},
  \bibinfo{person}{Lei Yan}, \bibinfo{person}{Teddy Seyed}, {and}
  \bibinfo{person}{Xing-Dong Yang}.} \bibinfo{year}{2019}\natexlab{}.
\newblock \showarticletitle{Tessutivo: Contextual Interactions on Interactive
  Fabrics with Inductive Sensing}. In \bibinfo{booktitle}{\emph{Proceedings of
  the 32nd Annual ACM Symposium on User Interface Software and Technology}}
  (New Orleans, LA, USA) \emph{(\bibinfo{series}{UIST '19})}.
  \bibinfo{publisher}{Association for Computing Machinery},
  \bibinfo{address}{New York, NY, USA}, \bibinfo{pages}{29–41}.
\newblock
\showISBNx{9781450368162}
\urldef\tempurl%
\url{https://doi.org/10.1145/3332165.3347897}
\showDOI{\tempurl}


\bibitem[Grimes et~al\mbox{.}(2002)]%
        {Grimes2002}
\bibfield{author}{\bibinfo{person}{C. Grimes}, \bibinfo{person}{C. Mungle},
  \bibinfo{person}{K. Zeng}, \bibinfo{person}{M. Jain}, \bibinfo{person}{W.
  Dreschel}, \bibinfo{person}{M. Paulose}, {and} \bibinfo{person}{K. Ong}.}
  \bibinfo{year}{2002}\natexlab{}.
\newblock \showarticletitle{Wireless Magnetoelastic Resonance Sensors: A
  Critical Review}.
\newblock \bibinfo{journal}{\emph{Sensors}} \bibinfo{volume}{2},
  \bibinfo{number}{7} (\bibinfo{date}{Jul} \bibinfo{year}{2002}),
  \bibinfo{pages}{294--313}.
\newblock
\urldef\tempurl%
\url{https://doi.org/10.3390/s20700294}
\showDOI{\tempurl}


\bibitem[Grosse-Puppendahl et~al\mbox{.}(2017)]%
        {Grosse-Puppendahl2017}
\bibfield{author}{\bibinfo{person}{Tobias Grosse-Puppendahl},
  \bibinfo{person}{Christian Holz}, \bibinfo{person}{Gabe Cohn},
  \bibinfo{person}{Raphael Wimmer}, \bibinfo{person}{Oskar Bechtold},
  \bibinfo{person}{Steve Hodges}, \bibinfo{person}{Matthew~S. Reynolds}, {and}
  \bibinfo{person}{Joshua~R. Smith}.} \bibinfo{year}{2017}\natexlab{}.
\newblock \showarticletitle{Finding Common Ground: A Survey of Capacitive
  Sensing in Human-Computer Interaction}. In
  \bibinfo{booktitle}{\emph{Proceedings of the 2017 CHI Conference on Human
  Factors in Computing Systems}} (Denver, Colorado, USA)
  \emph{(\bibinfo{series}{CHI '17})}. \bibinfo{publisher}{Association for
  Computing Machinery}, \bibinfo{address}{New York, NY, USA},
  \bibinfo{pages}{3293–3315}.
\newblock
\showISBNx{9781450346559}
\urldef\tempurl%
\url{https://doi.org/10.1145/3025453.3025808}
\showDOI{\tempurl}


\bibitem[Hajiaghajani et~al\mbox{.}(2021)]%
        {Hajiaghajani2021}
\bibfield{author}{\bibinfo{person}{A. Hajiaghajani}, \bibinfo{person}{A.H.
  Afandizadeh~Zargari}, \bibinfo{person}{M. Dautta}, \bibinfo{person}{A.
  Jimenez}, \bibinfo{person}{F. Kurdahi}, {and} \bibinfo{person}{P. Tseng}.}
  \bibinfo{year}{2021}\natexlab{}.
\newblock \showarticletitle{Textile-integrated metamaterials for near-field
  multibody area networks}.
\newblock \bibinfo{journal}{\emph{Nature Electronics}} \bibinfo{volume}{4},
  \bibinfo{number}{11} (\bibinfo{date}{Nov} \bibinfo{year}{2021}),
  \bibinfo{pages}{808--817}.
\newblock
\urldef\tempurl%
\url{https://doi.org/10.1038/s41928-021-00663-0}
\showDOI{\tempurl}


\bibitem[Hamdan et~al\mbox{.}(2018)]%
        {Hamdan2018}
\bibfield{author}{\bibinfo{person}{Nur Al-huda Hamdan}, \bibinfo{person}{Simon
  Voelker}, {and} \bibinfo{person}{Jan Borchers}.}
  \bibinfo{year}{2018}\natexlab{}.
\newblock \showarticletitle{Sketch\&Stitch: Interactive Embroidery for
  E-textiles}. In \bibinfo{booktitle}{\emph{Proceedings of the 2018 CHI
  Conference on Human Factors in Computing Systems}} (Montreal QC, Canada)
  \emph{(\bibinfo{series}{CHI '18})}. \bibinfo{publisher}{Association for
  Computing Machinery}, \bibinfo{address}{New York, NY, USA},
  \bibinfo{pages}{1–13}.
\newblock
\showISBNx{9781450356206}
\urldef\tempurl%
\url{https://doi.org/10.1145/3173574.3173656}
\showDOI{\tempurl}


\bibitem[Heller et~al\mbox{.}(2014)]%
        {Heller2014}
\bibfield{author}{\bibinfo{person}{Florian Heller}, \bibinfo{person}{Stefan
  Ivanov}, \bibinfo{person}{Chat Wacharamanotham}, {and} \bibinfo{person}{Jan
  Borchers}.} \bibinfo{year}{2014}\natexlab{}.
\newblock \showarticletitle{FabriTouch: exploring flexible touch input on
  textiles}. In \bibinfo{booktitle}{\emph{Proceedings of the 2014 ACM
  International Symposium on Wearable Computers}} (Seattle, Washington)
  \emph{(\bibinfo{series}{ISWC '14})}. \bibinfo{publisher}{Association for
  Computing Machinery}, \bibinfo{address}{New York, NY, USA},
  \bibinfo{pages}{59–62}.
\newblock
\showISBNx{9781450329699}
\urldef\tempurl%
\url{https://doi.org/10.1145/2634317.2634345}
\showDOI{\tempurl}


\bibitem[Huang et~al\mbox{.}(2021)]%
        {Huang2021}
\bibfield{author}{\bibinfo{person}{Kunpeng Huang}, \bibinfo{person}{Ruojia
  Sun}, \bibinfo{person}{Ximeng Zhang}, \bibinfo{person}{Md.~Tahmidul
  Islam~Molla}, \bibinfo{person}{Margaret Dunne}, \bibinfo{person}{Francois
  Guimbretiere}, {and} \bibinfo{person}{Cindy Hsin-Liu Kao}.}
  \bibinfo{year}{2021}\natexlab{}.
\newblock \showarticletitle{WovenProbe: Probing Possibilities for Weaving
  Fully-Integrated On-Skin Systems Deployable in the Field}. In
  \bibinfo{booktitle}{\emph{Proceedings of the 2021 ACM Designing Interactive
  Systems Conference}} (Virtual Event, USA) \emph{(\bibinfo{series}{DIS '21})}.
  \bibinfo{publisher}{Association for Computing Machinery},
  \bibinfo{address}{New York, NY, USA}, \bibinfo{pages}{1143–1158}.
\newblock
\showISBNx{9781450384766}
\urldef\tempurl%
\url{https://doi.org/10.1145/3461778.3462105}
\showDOI{\tempurl}


\bibitem[Hughes-Riley et~al\mbox{.}(2024)]%
        {10.1145/3613905.3648646}
\bibfield{author}{\bibinfo{person}{Theo Hughes-Riley},
  \bibinfo{person}{Arash~M. Shahidi}, \bibinfo{person}{Kalana Marasinghe},
  \bibinfo{person}{Zahra Rahemtulla}, \bibinfo{person}{Malindu Ehelagasthenna},
  \bibinfo{person}{Parvin Ebrahimi}, \bibinfo{person}{Richard Arm},
  \bibinfo{person}{Carlos Oliveira}, \bibinfo{person}{Lars~Erik Holmquist},
  {and} \bibinfo{person}{Tilak Dias}.} \bibinfo{year}{2024}\natexlab{}.
\newblock \showarticletitle{Wearable Electronic Textiles for Healthcare,
  Wellbeing, and Protective Applications}. In
  \bibinfo{booktitle}{\emph{Extended Abstracts of the CHI Conference on Human
  Factors in Computing Systems}} (Honolulu, HI, USA)
  \emph{(\bibinfo{series}{CHI EA '24})}. \bibinfo{publisher}{Association for
  Computing Machinery}, \bibinfo{address}{New York, NY, USA}, Article
  \bibinfo{articleno}{426}, \bibinfo{numpages}{5}~pages.
\newblock
\showISBNx{9798400703317}
\urldef\tempurl%
\url{https://doi.org/10.1145/3613905.3648646}
\showDOI{\tempurl}


\bibitem[Ibanez-Labiano and Alomainy(2020)]%
        {Ibanez-Labiano2020}
\bibfield{author}{\bibinfo{person}{I. Ibanez-Labiano} {and} \bibinfo{person}{A.
  Alomainy}.} \bibinfo{year}{2020}\natexlab{}.
\newblock \showarticletitle{Dielectric Characterization of Non-Conductive
  Fabrics for Temperature Sensing through Resonating Antenna Structures}.
\newblock \bibinfo{journal}{\emph{Materials}} \bibinfo{volume}{13},
  \bibinfo{number}{6} (\bibinfo{date}{Mar} \bibinfo{year}{2020}),
  \bibinfo{pages}{1271}.
\newblock
\urldef\tempurl%
\url{https://doi.org/10.3390/ma13061271}
\showDOI{\tempurl}


\bibitem[Jeon and Seo(2019)]%
        {Jeon2019}
\bibfield{author}{\bibinfo{person}{S.-J. Jeon} {and} \bibinfo{person}{D.-W.
  Seo}.} \bibinfo{year}{2019}\natexlab{}.
\newblock \showarticletitle{Coupling Coefficient Measurement Method with Simple
  Procedures Using a Two-Port Network Analyzer for a Multi-Coil WPT System}.
\newblock \bibinfo{journal}{\emph{Energies}} \bibinfo{volume}{12},
  \bibinfo{number}{20} (\bibinfo{date}{Oct} \bibinfo{year}{2019}),
  \bibinfo{pages}{3950}.
\newblock
\urldef\tempurl%
\url{https://doi.org/10.3390/en12203950}
\showDOI{\tempurl}


\bibitem[Kazemitabaar et~al\mbox{.}(2017)]%
        {Kazemitabaar2017}
\bibfield{author}{\bibinfo{person}{Majeed Kazemitabaar}, \bibinfo{person}{Jason
  McPeak}, \bibinfo{person}{Alexander Jiao}, \bibinfo{person}{Liang He},
  \bibinfo{person}{Thomas Outing}, {and} \bibinfo{person}{Jon~E. Froehlich}.}
  \bibinfo{year}{2017}\natexlab{}.
\newblock \showarticletitle{MakerWear: A Tangible Approach to Interactive
  Wearable Creation for Children}. In \bibinfo{booktitle}{\emph{Proceedings of
  the 2017 CHI Conference on Human Factors in Computing Systems}} (Denver,
  Colorado, USA) \emph{(\bibinfo{series}{CHI '17})}.
  \bibinfo{publisher}{Association for Computing Machinery},
  \bibinfo{address}{New York, NY, USA}, \bibinfo{pages}{133–145}.
\newblock
\showISBNx{9781450346559}
\urldef\tempurl%
\url{https://doi.org/10.1145/3025453.3025887}
\showDOI{\tempurl}


\bibitem[Kazemitabaar et~al\mbox{.}(2015)]%
        {Kazemitabaar2015}
\bibfield{author}{\bibinfo{person}{Majeed Kazemitabaar}, \bibinfo{person}{Leyla
  Norooz}, \bibinfo{person}{Mona~Leigh Guha}, {and} \bibinfo{person}{Jon~E.
  Froehlich}.} \bibinfo{year}{2015}\natexlab{}.
\newblock \showarticletitle{MakerShoe: towards a wearable e-textile
  construction kit to support creativity, playful making, and self-expression}.
  In \bibinfo{booktitle}{\emph{Proceedings of the 14th International Conference
  on Interaction Design and Children}} (Boston, Massachusetts)
  \emph{(\bibinfo{series}{IDC '15})}. \bibinfo{publisher}{Association for
  Computing Machinery}, \bibinfo{address}{New York, NY, USA},
  \bibinfo{pages}{449–452}.
\newblock
\showISBNx{9781450335904}
\urldef\tempurl%
\url{https://doi.org/10.1145/2771839.2771883}
\showDOI{\tempurl}


\bibitem[Klamka et~al\mbox{.}(2020)]%
        {Klamka2020}
\bibfield{author}{\bibinfo{person}{Konstantin Klamka}, \bibinfo{person}{Raimund
  Dachselt}, {and} \bibinfo{person}{J\"{u}rgen Steimle}.}
  \bibinfo{year}{2020}\natexlab{}.
\newblock \showarticletitle{Rapid Iron-On User Interfaces: Hands-on Fabrication
  of Interactive Textile Prototypes}. In \bibinfo{booktitle}{\emph{Proceedings
  of the 2020 CHI Conference on Human Factors in Computing Systems}} (Honolulu,
  HI, USA) \emph{(\bibinfo{series}{CHI '20})}. \bibinfo{publisher}{Association
  for Computing Machinery}, \bibinfo{address}{New York, NY, USA},
  \bibinfo{pages}{1–14}.
\newblock
\showISBNx{9781450367080}
\urldef\tempurl%
\url{https://doi.org/10.1145/3313831.3376220}
\showDOI{\tempurl}


\bibitem[Ku et~al\mbox{.}(2020)]%
        {Ku2020}
\bibfield{author}{\bibinfo{person}{Pin-Sung Ku}, \bibinfo{person}{Qijia Shao},
  \bibinfo{person}{Te-Yen Wu}, \bibinfo{person}{Jun Gong},
  \bibinfo{person}{Ziyan Zhu}, \bibinfo{person}{Xia Zhou}, {and}
  \bibinfo{person}{Xing-Dong Yang}.} \bibinfo{year}{2020}\natexlab{}.
\newblock \showarticletitle{ThreadSense: Locating Touch on an Extremely Thin
  Interactive Thread}. In \bibinfo{booktitle}{\emph{Proceedings of the 2020 CHI
  Conference on Human Factors in Computing Systems}} (Honolulu, HI, USA)
  \emph{(\bibinfo{series}{CHI '20})}. \bibinfo{publisher}{Association for
  Computing Machinery}, \bibinfo{address}{New York, NY, USA},
  \bibinfo{pages}{1–12}.
\newblock
\showISBNx{9781450367080}
\urldef\tempurl%
\url{https://doi.org/10.1145/3313831.3376779}
\showDOI{\tempurl}


\bibitem[Lara-Reyes et~al\mbox{.}(2022)]%
        {lara2022series}
\bibfield{author}{\bibinfo{person}{Josu{\'e} Lara-Reyes},
  \bibinfo{person}{Mario Ponce-Silva}, \bibinfo{person}{Leobardo
  Hern{\'a}ndez-Gonz{\'a}lez}, \bibinfo{person}{Susana~E DeLe{\'o}n-Aldaco},
  \bibinfo{person}{Claudia Cort{\'e}s-Garc{\'\i}a}, {and}
  \bibinfo{person}{Jazmin Ramirez-Hernandez}.} \bibinfo{year}{2022}\natexlab{}.
\newblock \showarticletitle{Series RLC Resonant Circuit Used as Frequency
  Multiplier}.
\newblock \bibinfo{journal}{\emph{Energies}} \bibinfo{volume}{15},
  \bibinfo{number}{24} (\bibinfo{year}{2022}), \bibinfo{pages}{9334}.
\newblock


\bibitem[Lee et~al\mbox{.}(2018)]%
        {Lee2018}
\bibfield{author}{\bibinfo{person}{K. Lee}, \bibinfo{person}{N. Chou}, {and}
  \bibinfo{person}{S. Kim}.} \bibinfo{year}{2018}\natexlab{}.
\newblock \showarticletitle{A Batteryless, Wireless Strain Sensor Using
  Resonant Frequency Modulation}.
\newblock \bibinfo{journal}{\emph{Sensors}} \bibinfo{volume}{18},
  \bibinfo{number}{11} (\bibinfo{date}{Nov} \bibinfo{year}{2018}),
  \bibinfo{pages}{3955}.
\newblock
\urldef\tempurl%
\url{https://doi.org/10.3390/s18113955}
\showDOI{\tempurl}


\bibitem[Lee and Chung(2009)]%
        {Lee2009}
\bibfield{author}{\bibinfo{person}{Y.-D. Lee} {and} \bibinfo{person}{W.-Y.
  Chung}.} \bibinfo{year}{2009}\natexlab{}.
\newblock \showarticletitle{Wireless sensor network based wearable smart shirt
  for ubiquitous health and activity monitoring}.
\newblock \bibinfo{journal}{\emph{Sensors and Actuators B: Chemical}}
  \bibinfo{volume}{140}, \bibinfo{number}{2} (\bibinfo{date}{Jul}
  \bibinfo{year}{2009}), \bibinfo{pages}{390--395}.
\newblock
\urldef\tempurl%
\url{https://doi.org/10.1016/j.snb.2009.04.040}
\showDOI{\tempurl}


\bibitem[Lin et~al\mbox{.}(2020)]%
        {Lin2020}
\bibfield{author}{\bibinfo{person}{R. Lin}, \bibinfo{person}{H.-J. Kim},
  \bibinfo{person}{S. Achavananthadith}, \bibinfo{person}{S.A. Kurt},
  \bibinfo{person}{S.C.C. Tan}, \bibinfo{person}{H. Yao},
  \bibinfo{person}{B.C.K. Tee}, \bibinfo{person}{J.K.W. Lee}, {and}
  \bibinfo{person}{J.S. Ho}.} \bibinfo{year}{2020}\natexlab{}.
\newblock \showarticletitle{Wireless battery-free body sensor networks using
  near-field-enabled clothing}.
\newblock \bibinfo{journal}{\emph{Nature Communications}} \bibinfo{volume}{11},
  \bibinfo{number}{1} (\bibinfo{date}{Jan} \bibinfo{year}{2020}),
  \bibinfo{pages}{444}.
\newblock
\urldef\tempurl%
\url{https://doi.org/10.1038/s41467-020-14311-2}
\showDOI{\tempurl}


\bibitem[Liu et~al\mbox{.}(2019)]%
        {Liu2019}
\bibfield{author}{\bibinfo{person}{Ruibo Liu}, \bibinfo{person}{Qijia Shao},
  \bibinfo{person}{Siqi Wang}, \bibinfo{person}{Christina Ru},
  \bibinfo{person}{Devin Balkcom}, {and} \bibinfo{person}{Xia Zhou}.}
  \bibinfo{year}{2019}\natexlab{}.
\newblock \showarticletitle{Reconstructing Human Joint Motion with
  Computational Fabrics}.
\newblock \bibinfo{journal}{\emph{Proc. ACM Interact. Mob. Wearable Ubiquitous
  Technol.}} \bibinfo{volume}{3}, \bibinfo{number}{1}, Article
  \bibinfo{articleno}{19} (\bibinfo{date}{March} \bibinfo{year}{2019}),
  \bibinfo{numpages}{26}~pages.
\newblock
\urldef\tempurl%
\url{https://doi.org/10.1145/3314406}
\showDOI{\tempurl}


\bibitem[Majumder et~al\mbox{.}(2017)]%
        {Majumder2017}
\bibfield{author}{\bibinfo{person}{S. Majumder}, \bibinfo{person}{T. Mondal},
  {and} \bibinfo{person}{M.J. Deen}.} \bibinfo{year}{2017}\natexlab{}.
\newblock \showarticletitle{Wearable Sensors for Remote Health Monitoring}.
\newblock \bibinfo{journal}{\emph{Sensors}} \bibinfo{volume}{17},
  \bibinfo{number}{1} (\bibinfo{date}{Jan} \bibinfo{year}{2017}),
  \bibinfo{pages}{130}.
\newblock
\urldef\tempurl%
\url{https://doi.org/10.3390/s17010130}
\showDOI{\tempurl}


\bibitem[Molla et~al\mbox{.}(2017)]%
        {Molla2017}
\bibfield{author}{\bibinfo{person}{Md. Tahmidul~Islam Molla},
  \bibinfo{person}{Steven Goodman}, \bibinfo{person}{Nicholas Schleif},
  \bibinfo{person}{Mary~Ellen Berglund}, \bibinfo{person}{Cade Zacharias},
  \bibinfo{person}{Crystal Compton}, {and} \bibinfo{person}{Lucy~E. Dunne}.}
  \bibinfo{year}{2017}\natexlab{}.
\newblock \showarticletitle{Surface-mount manufacturing for e-textile
  circuits}. In \bibinfo{booktitle}{\emph{Proceedings of the 2017 ACM
  International Symposium on Wearable Computers}} (Maui, Hawaii)
  \emph{(\bibinfo{series}{ISWC '17})}. \bibinfo{publisher}{Association for
  Computing Machinery}, \bibinfo{address}{New York, NY, USA},
  \bibinfo{pages}{18–25}.
\newblock
\showISBNx{9781450351881}
\urldef\tempurl%
\url{https://doi.org/10.1145/3123021.3123058}
\showDOI{\tempurl}


\bibitem[Ngai et~al\mbox{.}(2009)]%
        {Ngai2009}
\bibfield{author}{\bibinfo{person}{Grace Ngai}, \bibinfo{person}{Stephen~C.F.
  Chan}, \bibinfo{person}{Joey~C.Y. Cheung}, {and} \bibinfo{person}{Winnie~W.Y.
  Lau}.} \bibinfo{year}{2009}\natexlab{}.
\newblock \showarticletitle{The TeeBoard: an education-friendly construction
  platform for e-textiles and wearable computing}. In
  \bibinfo{booktitle}{\emph{Proceedings of the SIGCHI Conference on Human
  Factors in Computing Systems}} (Boston, MA, USA) \emph{(\bibinfo{series}{CHI
  '09})}. \bibinfo{publisher}{Association for Computing Machinery},
  \bibinfo{address}{New York, NY, USA}, \bibinfo{pages}{249–258}.
\newblock
\showISBNx{9781605582467}
\urldef\tempurl%
\url{https://doi.org/10.1145/1518701.1518742}
\showDOI{\tempurl}


\bibitem[Ngai et~al\mbox{.}(2010)]%
        {Ngai2010}
\bibfield{author}{\bibinfo{person}{Grace Ngai}, \bibinfo{person}{Stephen~C.F.
  Chan}, \bibinfo{person}{Vincent~T.Y. Ng}, \bibinfo{person}{Joey~C.Y. Cheung},
  \bibinfo{person}{Sam~S.S. Choy}, \bibinfo{person}{Winnie~W.Y. Lau}, {and}
  \bibinfo{person}{Jason~T.P. Tse}.} \bibinfo{year}{2010}\natexlab{}.
\newblock \showarticletitle{i*CATch: a scalable plug-n-play wearable computing
  framework for novices and children}. In \bibinfo{booktitle}{\emph{Proceedings
  of the SIGCHI Conference on Human Factors in Computing Systems}} (Atlanta,
  Georgia, USA) \emph{(\bibinfo{series}{CHI '10})}.
  \bibinfo{publisher}{Association for Computing Machinery},
  \bibinfo{address}{New York, NY, USA}, \bibinfo{pages}{443–452}.
\newblock
\showISBNx{9781605589299}
\urldef\tempurl%
\url{https://doi.org/10.1145/1753326.1753393}
\showDOI{\tempurl}


\bibitem[Olwal et~al\mbox{.}(2018)]%
        {Olwal2018}
\bibfield{author}{\bibinfo{person}{Alex Olwal}, \bibinfo{person}{Jon Moeller},
  \bibinfo{person}{Greg Priest-Dorman}, \bibinfo{person}{Thad Starner}, {and}
  \bibinfo{person}{Ben Carroll}.} \bibinfo{year}{2018}\natexlab{}.
\newblock \showarticletitle{I/O Braid: Scalable Touch-Sensitive Lighted Cords
  Using Spiraling, Repeating Sensing Textiles and Fiber Optics}. In
  \bibinfo{booktitle}{\emph{Proceedings of the 31st Annual ACM Symposium on
  User Interface Software and Technology}} (Berlin, Germany)
  \emph{(\bibinfo{series}{UIST '18})}. \bibinfo{publisher}{Association for
  Computing Machinery}, \bibinfo{address}{New York, NY, USA},
  \bibinfo{pages}{485–497}.
\newblock
\showISBNx{9781450359481}
\urldef\tempurl%
\url{https://doi.org/10.1145/3242587.3242638}
\showDOI{\tempurl}


\bibitem[Olwal et~al\mbox{.}(2020)]%
        {Olwal2020}
\bibfield{author}{\bibinfo{person}{Alex Olwal}, \bibinfo{person}{Thad Starner},
  {and} \bibinfo{person}{Gowa Mainini}.} \bibinfo{year}{2020}\natexlab{}.
\newblock \showarticletitle{E-Textile Microinteractions: Augmenting Twist with
  Flick, Slide and Grasp Gestures for Soft Electronics}. In
  \bibinfo{booktitle}{\emph{Proceedings of the 2020 CHI Conference on Human
  Factors in Computing Systems}} (Honolulu, HI, USA)
  \emph{(\bibinfo{series}{CHI '20})}. \bibinfo{publisher}{Association for
  Computing Machinery}, \bibinfo{address}{New York, NY, USA},
  \bibinfo{pages}{1–13}.
\newblock
\showISBNx{9781450367080}
\urldef\tempurl%
\url{https://doi.org/10.1145/3313831.3376236}
\showDOI{\tempurl}


\bibitem[Ossevoort(2013)]%
        {Ossevoort2013}
\bibfield{author}{\bibinfo{person}{S.H.W. Ossevoort}.}
  \bibinfo{year}{2013}\natexlab{}.
\newblock \bibinfo{booktitle}{\emph{Improving the sustainability of smart
  textiles}}.
\newblock \bibinfo{publisher}{Elsevier}, \bibinfo{pages}{399--419}.
\newblock


\bibitem[Parzer et~al\mbox{.}(2018)]%
        {Parzer2018}
\bibfield{author}{\bibinfo{person}{Patrick Parzer}, \bibinfo{person}{Florian
  Perteneder}, \bibinfo{person}{Kathrin Probst}, \bibinfo{person}{Christian
  Rendl}, \bibinfo{person}{Joanne Leong}, \bibinfo{person}{Sarah Schuetz},
  \bibinfo{person}{Anita Vogl}, \bibinfo{person}{Reinhard Schwoediauer},
  \bibinfo{person}{Martin Kaltenbrunner}, \bibinfo{person}{Siegfried Bauer},
  {and} \bibinfo{person}{Michael Haller}.} \bibinfo{year}{2018}\natexlab{}.
\newblock \showarticletitle{RESi: A Highly Flexible, Pressure-Sensitive,
  Imperceptible Textile Interface Based on Resistive Yarns}. In
  \bibinfo{booktitle}{\emph{Proceedings of the 31st Annual ACM Symposium on
  User Interface Software and Technology}} (Berlin, Germany)
  \emph{(\bibinfo{series}{UIST '18})}. \bibinfo{publisher}{Association for
  Computing Machinery}, \bibinfo{address}{New York, NY, USA},
  \bibinfo{pages}{745–756}.
\newblock
\showISBNx{9781450359481}
\urldef\tempurl%
\url{https://doi.org/10.1145/3242587.3242664}
\showDOI{\tempurl}


\bibitem[Parzer et~al\mbox{.}(2017)]%
        {Parzer2017}
\bibfield{author}{\bibinfo{person}{Patrick Parzer}, \bibinfo{person}{Adwait
  Sharma}, \bibinfo{person}{Anita Vogl}, \bibinfo{person}{J\"{u}rgen Steimle},
  \bibinfo{person}{Alex Olwal}, {and} \bibinfo{person}{Michael Haller}.}
  \bibinfo{year}{2017}\natexlab{}.
\newblock \showarticletitle{SmartSleeve: Real-time Sensing of Surface and
  Deformation Gestures on Flexible, Interactive Textiles, using a Hybrid
  Gesture Detection Pipeline}. In \bibinfo{booktitle}{\emph{Proceedings of the
  30th Annual ACM Symposium on User Interface Software and Technology}}
  (Qu\'{e}bec City, QC, Canada) \emph{(\bibinfo{series}{UIST '17})}.
  \bibinfo{publisher}{Association for Computing Machinery},
  \bibinfo{address}{New York, NY, USA}, \bibinfo{pages}{565–577}.
\newblock
\showISBNx{9781450349819}
\urldef\tempurl%
\url{https://doi.org/10.1145/3126594.3126652}
\showDOI{\tempurl}


\bibitem[Post et~al\mbox{.}(2000)]%
        {Post2000}
\bibfield{author}{\bibinfo{person}{E.R. Post}, \bibinfo{person}{M. Orth},
  \bibinfo{person}{P.R. Russo}, {and} \bibinfo{person}{N. Gershenfeld}.}
  \bibinfo{year}{2000}\natexlab{}.
\newblock \showarticletitle{E-broidery: design and fabrication of textile-based
  computing}.
\newblock \bibinfo{journal}{\emph{IBM Systems Journal}} \bibinfo{volume}{39},
  \bibinfo{number}{3-4} (\bibinfo{date}{Jul} \bibinfo{year}{2000}),
  \bibinfo{pages}{840--860}.
\newblock
\urldef\tempurl%
\url{https://doi.org/10.1147/sj.393.0840}
\showDOI{\tempurl}


\bibitem[Poupyrev et~al\mbox{.}(2016)]%
        {Poupyrev2016}
\bibfield{author}{\bibinfo{person}{Ivan Poupyrev}, \bibinfo{person}{Nan-Wei
  Gong}, \bibinfo{person}{Shiho Fukuhara}, \bibinfo{person}{Mustafa~Emre
  Karagozler}, \bibinfo{person}{Carsten Schwesig}, {and}
  \bibinfo{person}{Karen~E. Robinson}.} \bibinfo{year}{2016}\natexlab{}.
\newblock \showarticletitle{Project Jacquard: Interactive Digital Textiles at
  Scale}. In \bibinfo{booktitle}{\emph{Proceedings of the 2016 CHI Conference
  on Human Factors in Computing Systems}} (San Jose, California, USA)
  \emph{(\bibinfo{series}{CHI '16})}. \bibinfo{publisher}{Association for
  Computing Machinery}, \bibinfo{address}{New York, NY, USA},
  \bibinfo{pages}{4216–4227}.
\newblock
\showISBNx{9781450333627}
\urldef\tempurl%
\url{https://doi.org/10.1145/2858036.2858176}
\showDOI{\tempurl}


\bibitem[Riistama et~al\mbox{.}(2010)]%
        {Riistama2010}
\bibfield{author}{\bibinfo{person}{J. Riistama}, \bibinfo{person}{E.
  Aittokallio}, \bibinfo{person}{J. Verho}, {and} \bibinfo{person}{J.
  Lekkala}.} \bibinfo{year}{2010}\natexlab{}.
\newblock \showarticletitle{Totally passive wireless biopotential measurement
  sensor by utilizing inductively coupled resonance circuits}.
\newblock \bibinfo{journal}{\emph{Sensors and Actuators A: Physical}}
  \bibinfo{volume}{157}, \bibinfo{number}{2} (\bibinfo{date}{Feb}
  \bibinfo{year}{2010}), \bibinfo{pages}{313--321}.
\newblock
\urldef\tempurl%
\url{https://doi.org/10.1016/j.sna.2009.11.038}
\showDOI{\tempurl}


\bibitem[Sacristán-Riquelme et~al\mbox{.}(2009)]%
        {SACRISTANRIQUELME2009177}
\bibfield{author}{\bibinfo{person}{Jordi Sacristán-Riquelme},
  \bibinfo{person}{Fredy Segura-Quijano}, \bibinfo{person}{Antoni Baldi}, {and}
  \bibinfo{person}{M. {Teresa Osés}}.} \bibinfo{year}{2009}\natexlab{}.
\newblock \showarticletitle{Low power impedance measurement integrated circuit
  for sensor applications}.
\newblock \bibinfo{journal}{\emph{Microelectronics Journal}}
  \bibinfo{volume}{40}, \bibinfo{number}{1} (\bibinfo{year}{2009}),
  \bibinfo{pages}{177--184}.
\newblock
\showISSN{1879-2391}
\urldef\tempurl%
\url{https://doi.org/10.1016/j.mejo.2008.07.003}
\showDOI{\tempurl}


\bibitem[Saha et~al\mbox{.}(2018)]%
        {Saha2018}
\bibfield{author}{\bibinfo{person}{C. Saha}, \bibinfo{person}{I. Anya},
  \bibinfo{person}{C. Alexandru}, {and} \bibinfo{person}{R. Jinks}.}
  \bibinfo{year}{2018}\natexlab{}.
\newblock \showarticletitle{Wireless power transfer using relay resonators}.
\newblock \bibinfo{journal}{\emph{Applied Physics Letters}}
  \bibinfo{volume}{112}, \bibinfo{number}{26} (\bibinfo{date}{Jun}
  \bibinfo{year}{2018}), \bibinfo{pages}{263902}.
\newblock
\urldef\tempurl%
\url{https://doi.org/10.1063/1.5022032}
\showDOI{\tempurl}


\bibitem[Salpavaara(2018)]%
        {Salpavaara2018}
\bibfield{author}{\bibinfo{person}{T. Salpavaara}.}
  \bibinfo{year}{2018}\natexlab{}.
\newblock \emph{\bibinfo{title}{Inductively Coupled Passive Resonance Sensors:
  Readout Methods and Applications}}.
\newblock \bibinfo{thesistype}{Ph.\,D. Dissertation}. \bibinfo{school}{Tampere
  University of Technology}.
\newblock


\bibitem[Seo et~al\mbox{.}(2016)]%
        {Seo2016}
\bibfield{author}{\bibinfo{person}{D.-W. Seo}, \bibinfo{person}{J.-H. Lee},
  {and} \bibinfo{person}{H. Lee}.} \bibinfo{year}{2016}\natexlab{}.
\newblock \showarticletitle{A Study on Two-coil and Four-coil Wireless Power
  Transfer System Using Z-parameter Approach}.
\newblock \bibinfo{journal}{\emph{ETRI Journal}} (\bibinfo{date}{Feb}
  \bibinfo{year}{2016}).
\newblock
\urldef\tempurl%
\url{https://doi.org/10.4218/etrij.16.0115.0692}
\showDOI{\tempurl}


\bibitem[Seyed et~al\mbox{.}(2021)]%
        {Seyed2021}
\bibfield{author}{\bibinfo{person}{Teddy Seyed}, \bibinfo{person}{James
  Devine}, \bibinfo{person}{Joe Finney}, \bibinfo{person}{Michal Moskal},
  \bibinfo{person}{Peli de Halleux}, \bibinfo{person}{Steve Hodges},
  \bibinfo{person}{Thomas Ball}, {and} \bibinfo{person}{Asta Roseway}.}
  \bibinfo{year}{2021}\natexlab{}.
\newblock \showarticletitle{Rethinking the Runway: Using Avant-Garde Fashion To
  Design a System for Wearables}. In \bibinfo{booktitle}{\emph{Proceedings of
  the 2021 CHI Conference on Human Factors in Computing Systems}} (Yokohama,
  Japan) \emph{(\bibinfo{series}{CHI '21})}. \bibinfo{publisher}{Association
  for Computing Machinery}, \bibinfo{address}{New York, NY, USA}, Article
  \bibinfo{articleno}{45}, \bibinfo{numpages}{15}~pages.
\newblock
\showISBNx{9781450380966}
\urldef\tempurl%
\url{https://doi.org/10.1145/3411764.3445643}
\showDOI{\tempurl}


\bibitem[Sun et~al\mbox{.}(2022)]%
        {Sun2022}
\bibfield{author}{\bibinfo{person}{X. Sun}, \bibinfo{person}{H. Wu},
  \bibinfo{person}{X. Tan}, \bibinfo{person}{W. Wang}, \bibinfo{person}{L. Ye},
  {and} \bibinfo{person}{K. Song}.} \bibinfo{year}{2022}\natexlab{}.
\newblock \showarticletitle{Frequency-Modulated Signal Measurement Using
  Closed-Loop Methodology}.
\newblock \bibinfo{journal}{\emph{Sensors}} \bibinfo{volume}{22},
  \bibinfo{number}{20} (\bibinfo{date}{Oct} \bibinfo{year}{2022}),
  \bibinfo{pages}{7822}.
\newblock
\urldef\tempurl%
\url{https://doi.org/10.3390/s22207822}
\showDOI{\tempurl}


\bibitem[Sun et~al\mbox{.}(2021)]%
        {Sun2021}
\bibfield{author}{\bibinfo{person}{Z. Sun}, \bibinfo{person}{H. Fang},
  \bibinfo{person}{B. Xu}, \bibinfo{person}{L. Yang}, \bibinfo{person}{H. Niu},
  \bibinfo{person}{H. Wang}, \bibinfo{person}{D. Chen}, \bibinfo{person}{Y.
  Liu}, \bibinfo{person}{Z. Wang}, \bibinfo{person}{Y. Wang}, {and}
  \bibinfo{person}{Q. Guo}.} \bibinfo{year}{2021}\natexlab{}.
\newblock \showarticletitle{Flexible Wireless Passive LC Pressure Sensor with
  Design Methodology and Cost-Effective Preparation}.
\newblock \bibinfo{journal}{\emph{Micromachines}} \bibinfo{volume}{12},
  \bibinfo{number}{8} (\bibinfo{date}{Aug} \bibinfo{year}{2021}),
  \bibinfo{pages}{976}.
\newblock
\urldef\tempurl%
\url{https://doi.org/10.3390/mi12080976}
\showDOI{\tempurl}


\bibitem[Tat et~al\mbox{.}(2022)]%
        {Tat2022}
\bibfield{author}{\bibinfo{person}{T. Tat}, \bibinfo{person}{G. Chen},
  \bibinfo{person}{X. Zhao}, \bibinfo{person}{Y. Zhou}, \bibinfo{person}{J.
  Xu}, {and} \bibinfo{person}{J. Chen}.} \bibinfo{year}{2022}\natexlab{}.
\newblock \showarticletitle{Smart Textiles for Healthcare and Sustainability}.
\newblock \bibinfo{journal}{\emph{ACS Nano}} \bibinfo{volume}{16},
  \bibinfo{number}{9} (\bibinfo{date}{Sep} \bibinfo{year}{2022}),
  \bibinfo{pages}{13301--13313}.
\newblock
\urldef\tempurl%
\url{https://doi.org/10.1021/acsnano.2c06287}
\showDOI{\tempurl}


\bibitem[Timmins(2009)]%
        {Timmins2009}
\bibfield{author}{\bibinfo{person}{M. Timmins}.}
  \bibinfo{year}{2009}\natexlab{}.
\newblock \bibinfo{booktitle}{\emph{Environmental and waste issues concerning
  the production of smart clothes and wearable technology}}.
\newblock \bibinfo{publisher}{Woodhead Publishing}, \bibinfo{pages}{319--331}.
\newblock


\bibitem[Vogl et~al\mbox{.}(2017)]%
        {Vogl2017}
\bibfield{author}{\bibinfo{person}{Anita Vogl}, \bibinfo{person}{Patrick
  Parzer}, \bibinfo{person}{Teo Babic}, \bibinfo{person}{Joanne Leong},
  \bibinfo{person}{Alex Olwal}, {and} \bibinfo{person}{Michael Haller}.}
  \bibinfo{year}{2017}\natexlab{}.
\newblock \showarticletitle{StretchEBand: Enabling Fabric-based Interactions
  through Rapid Fabrication of Textile Stretch Sensors}. In
  \bibinfo{booktitle}{\emph{Proceedings of the 2017 CHI Conference on Human
  Factors in Computing Systems}} (Denver, Colorado, USA)
  \emph{(\bibinfo{series}{CHI '17})}. \bibinfo{publisher}{Association for
  Computing Machinery}, \bibinfo{address}{New York, NY, USA},
  \bibinfo{pages}{2617–2627}.
\newblock
\showISBNx{9781450346559}
\urldef\tempurl%
\url{https://doi.org/10.1145/3025453.3025938}
\showDOI{\tempurl}


\bibitem[Wagih et~al\mbox{.}(2020)]%
        {Wagih2020}
\bibfield{author}{\bibinfo{person}{M. Wagih}, \bibinfo{person}{A. Komolafe},
  {and} \bibinfo{person}{B. Zaghari}.} \bibinfo{year}{2020}\natexlab{}.
\newblock \showarticletitle{Dual-Receiver Wearable 6.78 MHz Resonant Inductive
  Wireless Power Transfer Glove Using Embroidered Textile Coils}.
\newblock \bibinfo{journal}{\emph{IEEE Access}}  \bibinfo{volume}{8}
  (\bibinfo{year}{2020}), \bibinfo{pages}{24630--24642}.
\newblock


\bibitem[Wang et~al\mbox{.}(2019)]%
        {Flextouch}
\bibfield{author}{\bibinfo{person}{Yuntao Wang}, \bibinfo{person}{Jianyu Zhou},
  \bibinfo{person}{Hanchuan Li}, \bibinfo{person}{Tengxiang Zhang},
  \bibinfo{person}{Minxuan Gao}, \bibinfo{person}{Zhuolin Cheng},
  \bibinfo{person}{Chun Yu}, \bibinfo{person}{Shwetak Patel}, {and}
  \bibinfo{person}{Yuanchun Shi}.} \bibinfo{year}{2019}\natexlab{}.
\newblock \showarticletitle{FlexTouch: Enabling Large-Scale Interaction Sensing
  Beyond Touchscreens Using Flexible and Conductive Materials}.
\newblock \bibinfo{journal}{\emph{Proc. ACM Interact. Mob. Wearable Ubiquitous
  Technol.}} \bibinfo{volume}{3}, \bibinfo{number}{3}, Article
  \bibinfo{articleno}{109} (\bibinfo{date}{Sept.} \bibinfo{year}{2019}),
  \bibinfo{numpages}{20}~pages.
\newblock
\urldef\tempurl%
\url{https://doi.org/10.1145/3351267}
\showDOI{\tempurl}


\bibitem[Wang(2013)]%
        {Wang2013}
\bibfield{author}{\bibinfo{person}{Z.L. Wang}.}
  \bibinfo{year}{2013}\natexlab{}.
\newblock \showarticletitle{Triboelectric Nanogenerators as New Energy
  Technology for Self-Powered Systems and as Active Mechanical and Chemical
  Sensors}.
\newblock \bibinfo{journal}{\emph{ACS Nano}} \bibinfo{volume}{7},
  \bibinfo{number}{11} (\bibinfo{date}{Nov} \bibinfo{year}{2013}),
  \bibinfo{pages}{9533--9557}.
\newblock
\urldef\tempurl%
\url{https://doi.org/10.1021/nn404614z}
\showDOI{\tempurl}


\bibitem[Wu et~al\mbox{.}(2020a)]%
        {Wu2020Fabriccio}
\bibfield{author}{\bibinfo{person}{Te-Yen Wu}, \bibinfo{person}{Shutong Qi},
  \bibinfo{person}{Junchi Chen}, \bibinfo{person}{MuJie Shang},
  \bibinfo{person}{Jun Gong}, \bibinfo{person}{Teddy Seyed}, {and}
  \bibinfo{person}{Xing-Dong Yang}.} \bibinfo{year}{2020}\natexlab{a}.
\newblock \showarticletitle{Fabriccio: Touchless Gestural Input on Interactive
  Fabrics}. In \bibinfo{booktitle}{\emph{Proceedings of the 2020 CHI Conference
  on Human Factors in Computing Systems}} (Honolulu, HI, USA)
  \emph{(\bibinfo{series}{CHI '20})}. \bibinfo{publisher}{Association for
  Computing Machinery}, \bibinfo{address}{New York, NY, USA},
  \bibinfo{pages}{1–14}.
\newblock
\showISBNx{9781450367080}
\urldef\tempurl%
\url{https://doi.org/10.1145/3313831.3376681}
\showDOI{\tempurl}


\bibitem[Wu et~al\mbox{.}(2020b)]%
        {Wu2020Capacitivo}
\bibfield{author}{\bibinfo{person}{Te-Yen Wu}, \bibinfo{person}{Lu Tan},
  \bibinfo{person}{Yuji Zhang}, \bibinfo{person}{Teddy Seyed}, {and}
  \bibinfo{person}{Xing-Dong Yang}.} \bibinfo{year}{2020}\natexlab{b}.
\newblock \showarticletitle{Capacitivo: Contact-Based Object Recognition on
  Interactive Fabrics using Capacitive Sensing}. In
  \bibinfo{booktitle}{\emph{Proceedings of the 33rd Annual ACM Symposium on
  User Interface Software and Technology}} (Virtual Event, USA)
  \emph{(\bibinfo{series}{UIST '20})}. \bibinfo{publisher}{Association for
  Computing Machinery}, \bibinfo{address}{New York, NY, USA},
  \bibinfo{pages}{649–661}.
\newblock
\showISBNx{9781450375146}
\urldef\tempurl%
\url{https://doi.org/10.1145/3379337.3415829}
\showDOI{\tempurl}


\bibitem[Wu et~al\mbox{.}(2021)]%
        {Wu2021}
\bibfield{author}{\bibinfo{person}{Te-Yen Wu}, \bibinfo{person}{Zheer Xu},
  \bibinfo{person}{Xing-Dong Yang}, \bibinfo{person}{Steve Hodges}, {and}
  \bibinfo{person}{Teddy Seyed}.} \bibinfo{year}{2021}\natexlab{}.
\newblock \showarticletitle{Project Tasca: Enabling Touch and Contextual
  Interactions with a Pocket-based Textile Sensor}. In
  \bibinfo{booktitle}{\emph{Proceedings of the 2021 CHI Conference on Human
  Factors in Computing Systems}} (Yokohama, Japan) \emph{(\bibinfo{series}{CHI
  '21})}. \bibinfo{publisher}{Association for Computing Machinery},
  \bibinfo{address}{New York, NY, USA}, Article \bibinfo{articleno}{4},
  \bibinfo{numpages}{13}~pages.
\newblock
\showISBNx{9781450380966}
\urldef\tempurl%
\url{https://doi.org/10.1145/3411764.3445712}
\showDOI{\tempurl}


\bibitem[Xu et~al\mbox{.}(2013)]%
        {Xu2013}
\bibfield{author}{\bibinfo{person}{W. Xu}, \bibinfo{person}{M.-C. Huang},
  \bibinfo{person}{N. Amini}, \bibinfo{person}{L. He}, {and}
  \bibinfo{person}{M. Sarrafzadeh}.} \bibinfo{year}{2013}\natexlab{}.
\newblock \showarticletitle{eCushion: A Textile Pressure Sensor Array Design
  and Calibration for Sitting Posture Analysis}.
\newblock \bibinfo{journal}{\emph{IEEE Sensors Journal}} \bibinfo{volume}{13},
  \bibinfo{number}{10} (\bibinfo{date}{Oct} \bibinfo{year}{2013}),
  \bibinfo{pages}{3926--3934}.
\newblock
\urldef\tempurl%
\url{https://doi.org/10.1109/JSEN.2013.2259589}
\showDOI{\tempurl}


\bibitem[Yang et~al\mbox{.}(2024)]%
        {Singlebodycoupledfiber}
\bibfield{author}{\bibinfo{person}{Weifeng Yang}, \bibinfo{person}{Shaomei
  Lin}, \bibinfo{person}{Wei Gong}, \bibinfo{person}{Rongzhou Lin},
  \bibinfo{person}{Chengmei Jiang}, \bibinfo{person}{Xin Yang},
  \bibinfo{person}{Yunhao Hu}, \bibinfo{person}{Jingjie Wang},
  \bibinfo{person}{Xiao Xiao}, \bibinfo{person}{Kerui Li},
  \bibinfo{person}{Yaogang Li}, \bibinfo{person}{Qinghong Zhang},
  \bibinfo{person}{John~S. Ho}, \bibinfo{person}{Yuxin Liu},
  \bibinfo{person}{Chengyi Hou}, {and} \bibinfo{person}{Hongzhi Wang}.}
  \bibinfo{year}{2024}\natexlab{}.
\newblock \showarticletitle{Single body-coupled fiber enables chipless textile
  electronics}.
\newblock \bibinfo{journal}{\emph{Science}} \bibinfo{volume}{384},
  \bibinfo{number}{6691} (\bibinfo{year}{2024}), \bibinfo{pages}{74--81}.
\newblock
\urldef\tempurl%
\url{https://doi.org/10.1126/science.adk3755}
\showDOI{\tempurl}
\showeprint{https://www.science.org/doi/pdf/10.1126/science.adk3755}


\bibitem[Ye et~al\mbox{.}(2022)]%
        {Ye2022}
\bibfield{author}{\bibinfo{person}{Huizhong Ye}, \bibinfo{person}{Chi-Jung
  Lee}, \bibinfo{person}{Te-Yen Wu}, \bibinfo{person}{Xing-Dong Yang},
  \bibinfo{person}{Bing-Yu Chen}, {and} \bibinfo{person}{Rong-Hao Liang}.}
  \bibinfo{year}{2022}\natexlab{}.
\newblock \showarticletitle{Body-Centric NFC: Body-Centric Interaction with NFC
  Devices Through Near-Field Enabled Clothing}. In
  \bibinfo{booktitle}{\emph{Proceedings of the 2022 ACM Designing Interactive
  Systems Conference}} (Virtual Event, Australia) \emph{(\bibinfo{series}{DIS
  '22})}. \bibinfo{publisher}{Association for Computing Machinery},
  \bibinfo{address}{New York, NY, USA}, \bibinfo{pages}{1626–1639}.
\newblock
\showISBNx{9781450393584}
\urldef\tempurl%
\url{https://doi.org/10.1145/3532106.3534569}
\showDOI{\tempurl}


\bibitem[Yu et~al\mbox{.}(2024)]%
        {yu2024seamposerepurposingseamscapacitive}
\bibfield{author}{\bibinfo{person}{Tianhong~Catherine Yu},
  \bibinfo{person}{Manru~Mary Zhang}, \bibinfo{person}{Peter He},
  \bibinfo{person}{Chi-Jung Lee}, \bibinfo{person}{Cassidy Cheesman},
  \bibinfo{person}{Saif Mahmud}, \bibinfo{person}{Ruidong Zhang},
  \bibinfo{person}{Francois Guimbretiere}, {and} \bibinfo{person}{Cheng
  Zhang}.} \bibinfo{year}{2024}\natexlab{}.
\newblock \showarticletitle{SeamPose: Repurposing Seams as Capacitive Sensors
  in a Shirt for Upper-Body Pose Tracking}. In
  \bibinfo{booktitle}{\emph{Proceedings of the 37th Annual ACM Symposium on
  User Interface Software and Technology}} (Pittsburgh, PA, USA)
  \emph{(\bibinfo{series}{UIST '24})}. \bibinfo{publisher}{Association for
  Computing Machinery}, \bibinfo{address}{New York, NY, USA}, Article
  \bibinfo{articleno}{72}, \bibinfo{numpages}{13}~pages.
\newblock
\showISBNx{9798400706288}
\urldef\tempurl%
\url{https://doi.org/10.1145/3654777.3676341}
\showDOI{\tempurl}


\bibitem[Zhu et~al\mbox{.}(2015)]%
        {Zhu2015}
\bibfield{author}{\bibinfo{person}{D. Zhu}, \bibinfo{person}{L. Clare},
  \bibinfo{person}{B.H. Stark}, {and} \bibinfo{person}{S.P. Beeby}.}
  \bibinfo{year}{2015}\natexlab{}.
\newblock \showarticletitle{Near field wireless power transfer using curved
  relay resonators for extended transfer distance}.
\newblock \bibinfo{journal}{\emph{Journal of Physics: Conference Series}}
  \bibinfo{volume}{660} (\bibinfo{date}{Dec} \bibinfo{year}{2015}),
  \bibinfo{pages}{012136}.
\newblock
\urldef\tempurl%
\url{https://doi.org/10.1088/1742-6596/660/1/012136}
\showDOI{\tempurl}


\bibitem[Zhu and Kao(2022)]%
        {Zhu2022}
\bibfield{author}{\bibinfo{person}{Jingwen Zhu} {and}
  \bibinfo{person}{Hsin-Liu~(Cindy) Kao}.} \bibinfo{year}{2022}\natexlab{}.
\newblock \showarticletitle{Scaling E-Textile Production: Understanding the
  Challenges of Soft Wearable Production for Individual Creators}. In
  \bibinfo{booktitle}{\emph{Proceedings of the 2022 ACM International Symposium
  on Wearable Computers}} (Cambridge, United Kingdom)
  \emph{(\bibinfo{series}{ISWC '22})}. \bibinfo{publisher}{Association for
  Computing Machinery}, \bibinfo{address}{New York, NY, USA},
  \bibinfo{pages}{94–99}.
\newblock
\showISBNx{9781450394246}
\urldef\tempurl%
\url{https://doi.org/10.1145/3544794.3558475}
\showDOI{\tempurl}


\end{thebibliography}

\appendix
\section{Impedance Approximation at Resonant Frequency of Sensor Circuit}
\label{sec_appendix_1}
\MK{In the sensor value estimation algorithm, we approximate the system impedance to Eq \ref{eq_rf} when a sensor circuit is at its resonant frequency and has low resistance.
In this section, we provide the details of the equation's derivation and explain why the resonant frequencies of sensor circuits are not at the lowest peak in the impedance spectrum.}

\MK{
Based on the equivalent circuit model in Section \ref{sec_equivalent_circuit}, the reciprocal of the impedance of $i^{th}$ sensor circuit ($Z_i(f)$) can be described as the following equation:
}

\begin{gather}
\frac{1}{Z_i(f)} =\frac{1}{\frac{1}{\frac{1}{(2\pi fl_i-\frac{1}{2\pi fc_i})j + r_i}+2\pi fC_{line_i}j}+2\pi fL_{line_i}j+R_{line_i}}
\end{gather}

\MK{
When the frequency ($f$) reaches the resonant frequency of $k^{th}$ sensor circuit ($2\pi f=\frac{1}{\sqrt{l_kc_k}}$), the reciprocal of $Z_i(f)\ (i=k)$ becomes the following equation:
}

\begin{gather}
    \frac{1}{Z_k(f)} = \frac{1}{\frac{1}{2\pi f C_{line_k}j+\frac{1}{r_k}}+2\pi f L_{line_k}j + R_{line_k}}
\end{gather}

\MK{
If the transmission lines' capacitance ($C_{line_k}$) is small enough, the capacitive reactance of the transmission lines becomes negligible compared to the resistance of the sensor circuit. In this case, the resistance of the sensor circuit ($r_k$) dominates the overall impedance. The equation can then be approximated as:
}

\begin{gather}
    \frac{1}{Z_k(f)} = \frac{1}{\frac{1}{2\pi f C_{line_k}j+\frac{1}{r_k}}+2\pi f L_{line_k}j + R_{line_k}} \approx \frac{1}{r_k + R_{line_k} + 2\pi fL_{line_k}j}
\end{gather}

\MK{
Next, we calculate the total impedance of the smart textile interface using Eq \ref{eq_zs}. We add up values of the reciprocal of impedance for all sensor circuits, denoted as $\Sigma_1^n\frac{1}{Z_i(f)}$.
Given the assumption that resonant frequencies of each sensor circuit are separated with enough frequency gaps (or in other word, when one sensor circuit reaches its resonant point, the frequency is far away from other sensor circuits' resonant points), we can calculate $\Sigma_1^n\frac{1}{Z_i(f)}$ at the resonant frequency of $k^{th}$ sensor circuit as follows:
}

\begin{gather}
    \Sigma_1^n\frac{1}{Z_i(f)} = \frac{1}{r_k + R_{line_k} + 2\pi fL_{line_k}j} + \Sigma_{i\neq k} \frac{1}{Z_i(f)}
\end{gather}

\MK{
As frequency ($f$) is away from the resonant frequency of $i^{th}\ (i \neq k)$ sensor circuit, the impedance of $i^{th}$ sensor circuit is high enough and $\frac{1}{Z_i(f)}$ can be calculated as:
}

\begin{gather}
    \frac{1}{Z_i(f)} = \frac{1}{\frac{1}{2\pi f C_{line_i}j} + 2\pi fL_{line_i}j + R_{line_i}} \ (i \neq k)
\end{gather}

\MK{
When the transmission line is not too long, the frequency of the resonance formed by line capacitance ${C_{line_i}}$ and inductance ($L_{line_i}$) is much higher than the resonant frequency of $k^{th}$ sensor circuit (for example, 40cm twisted transmission line owns a resonant frequency at 112MHz) and $R_{line_i}$ is small, resulting a small value of $\frac{1}{Z_i(f)}$ within frequency range ($\lvert\frac{1}{Z_i(f)}\rvert$ $\leq$ 0.004 when $f\leq$ 30MHz). Then, assuming that the capacitive or inductive sensor circuit is with low resistance (e.g. $r_k + R_{line_k}$ $\le$ 15), the reciprocal of the impedance of $k^{th}$ sensor circuit ($\frac{1}{Z_k(f)}$) is much higher than the rest sensor circuits ($\lvert\frac{1}{Z_k(f)}\rvert$ $\geq$ 0.041 when $f\leq$ 30MHz and line length = 40cm), making influence of other sensor circuits negligible.
}

\MK{
Then, by combining Eq \ref{eq_z} to Eq \ref{eq_zs} and the approximation result, we can calculate $Z(f)$ as:
}

\begin{gather}
    Z(f) = \frac{1}{\frac{1}{(2\pi fL_t)j-\frac{((2\pi f)k\sqrt{(L_t L_r)}j)^2}{r_k + R_{line_k} + 2\pi f(L_{line_i}j+L_r)j}}+2\pi fC_{SMA}j} 
\end{gather}

\MK{
Subsequently, $r_i$ and $R_{line_k}$ can be neglected due to the dominance of transmission line and receiver coil' s inductance. For example, at 7 MHz, the inductive impedance is approximately $200\Omega$ without $r_i$ and $R_{line_k}$, which is close to the absolute impedance ($\sqrt{(200\Omega)^2 + (15\Omega)^2} \approx 200.5\Omega$) when the sum of $r_i$ and $R_{line_k}$ is $15\Omega$. Thus, we can approximate the calculation of $Z(f)$ (in terms of magnitude) by neglecting $r_k$ and $R_{line_k}$, finally gaining a result as follows:
}

\begin{gather}
    \label{eq_approximation_z_final}
    \lvert Z(f)\rvert = \lvert\frac{1}{\frac{1}{(2\pi fL_t)j-\frac{((2\pi f)k\sqrt{(L_t L_r)}j)^2}{2\pi f(L_{line_i}j+L_r)j}}+2\pi fC_{SMA}j}\rvert 
\end{gather}

\MK{
Give this, higher resistance in the sensor circuit may reduce this approximation's accuracy, but its precision improves as frequency increases. Simulation results in Section \ref{sec_simulation_based_evaluation} also support these findings.
}

\MK{
Finally,  Eq. \ref{eq_approximation_z_final} also explains why the impedance spectrum lowest peaks deviate from the resonant frequencies of sensor circuits. These deviations result from combined factors, including receiver coil inductance and transmission line's impedance. Additionally, if the resonant frequencies of the sensor circuits are too close to each other, their interactions can cause further shifts in the impedance spectrum.
}
\end{document}